\documentclass[a4paper,11pt]{article}
\pdfoutput=1 % if your are submitting a pdflatex (i.e. if you have
\usepackage{jcappub} % for details on the use of the package, please
\usepackage{indentfirst}
\usepackage{appendix}
\usepackage{multirow}

\usepackage{array}
\usepackage{enumerate}
\usepackage{graphicx}
\usepackage{dcolumn}
\usepackage{bm}
\usepackage{amssymb}
\usepackage{latexsym}
\usepackage{booktabs}
\usepackage{amsmath}
\usepackage{multirow}
\usepackage{url}

\begin{document}

\title{Prospects for measuring dark energy with 21 cm intensity mapping experiments}

\author[a]{Peng-Ju Wu}

\author[a,b,1]{and Xin Zhang\note{Corresponding author}}

\affiliation[a]{Department of Physics, College of Sciences, Northeastern
University, Shenyang 110819, China}
\affiliation[b]{MOE Key Laboratory of Data Analytics and Optimization for Smart Industry, Northeastern University, Shenyang 110819, China}

\emailAdd{wupengju@stumail.neu.edu.cn, zhangxin@mail.neu.edu.cn}

\abstract{
Using the 21 cm intensity mapping (IM) technique can efficiently perform large-scale neutral hydrogen (\textsc{H\,i}) surveys, and this method has great potential for measuring dark-energy parameters. Some 21 cm IM experiments aiming at measuring dark energy in the redshift range of $0<z<3$ have been proposed and performed, in which the typical ones using single-dish mode include e.g., BINGO, FAST, and SKA1-MID, and those using interferometric mode include e.g., HIRAX, CHIME, and Tianlai. In this work, we make a forecast for these typical 21 cm IM experiments on their capability of measuring parameters of dark energy. We find that the interferometers have great advantages in constraining cosmological parameters. In particular, the Tianlai cylinder array alone can achieve the standard of precision cosmology for the $\Lambda$CDM model (i.e., the precision of parameters is better than 1\%). However, for constraining dynamical dark energy, we find that SKA1-MID performs very well. We show that the simulated 21 cm IM data can break the parameter degeneracies inherent in the CMB data, and CMB+SKA1-MID offers $\sigma(w)=0.013$ in the $w$CDM model, and $\sigma(w_0)=0.080$ and $\sigma(w_a)=0.25$ in the CPL model. Compared with CMB+BAO+SN, Tianlai can provide tighter constraints in $\Lambda$CDM and $w$CDM, but looser constraints (tighter than CMB+BAO) in CPL, and the combination CMB+BAO+SN+Tianlai gives $\sigma(w)=0.013$, $\sigma(w_0)=0.055$, and $\sigma(w_a)=0.13$. In addition, it is found that the synergy of FAST ($0<z<0.35$)+SKA1-MID ($0.35<z<0.77$)+Tianlai ($0.77<z<2.55$) offers a very promising survey strategy. Finally, we find that the residual foreground contamination amplitude has a considerable impact on constraint results. We show that in the future 21 cm IM experiments will provide a powerful probe for exploring the nature of dark energy.}

\maketitle
\section{Introduction}\label{sec1}

Type Ia supernovae (SNe Ia) observations have revealed that the expansion of the universe is accelerating \citep{Perlmutter:1998np, Riess:1998cb}, indicating that the gravitational force dominating the universe is repulsive rather than attractive. The reason for this acceleration remains unknown, but it has been proposed that it is dark energy that causes the cosmic expansion rate to increase because dark energy yields repulsive gravity. In order to study the nature of dark energy, it is necessary to precisely measure the late-time expansion history of the universe \citep{Weinberg:2013agg}. The baryon acoustic oscillations (BAO), as a rather useful observational tool, has been playing an important role in this field. They are remnant ripples in the distribution of galaxies that are imprinted by primordial sound waves in the photon-baryon plasma prior to the recombination epoch. The BAO scale, which is equal to the sound speed times the age of the universe at the decoupling, provides a standard ruler to measure the angular diameter distance $D_{\rm A}(z)$ and the Hubble parameter $H(z)$ \citep{Blake:2003rh, Seo:2003pu}, and hence allows accurate measurements of the expansion history of the universe. In addition to observing the expansion history through the BAO features in the matter power spectrum, the large-scale structures (LSS) measurements can also provide the structure growth rate $f(z)$ from redshift space distortions (RSDs). The growth rate works as a non-geometric probe of gravity over cosmic time, and thus has the potential to distinguish between dynamical dark energy and modified gravity (e.g., Refs. \citep{Guzzo:2008ac, Wang:2007ht, Beutler:2013yhm, Samushia:2013yga, Li:2015poa, Zhao:2017jma}).

The BAO scale has been measured using galaxy surveys and Ly$\alpha$ forest measurements to map the distribution of matter (e.g., Refs.~\citep{Beutler:2011hx, Anderson:2013zyy, Delubac:2014aqe, Ross:2014qpa, Alam:2016hwk}). This is a long process that requires resolving individual galaxies, and the redshift is limited in scope. However, the BAO scale can also be measured through the 21 cm emission from neutral hydrogen (\textsc{H\,i}). In the post-reionization era ($z\lesssim6$), most of the \textsc{H\,i} is thought to exist in dense gas clouds embedded in galaxies, where it is shielded from ionizing photons. Therefore, \textsc{H\,i} is a tracer of matter on cosmological scales. Theoretically, we could do cosmology by observing enough \textsc{H\,i}-emitting galaxies, but in practice, it will be a very tough task consuming a lot of observing time. Instead, we can map the intensity of 21 cm signal at a resolution much lower than that of individual galaxies, but it is still high enough to measure the BAO scale. This technique with much faster sky survey speed, known as the 21 cm intensity mapping (IM), provides us with an efficient way to measure the large-scale matter power spectrum, from which the BAO and RSD signals can be extracted (see e.g., Refs.~\citep{McQuinn:2005hk, Loeb:2008hg, Mao:2008ug, Lidz:2011dx, Battye:2012tg, Xu:2014bya, Bull:2014rha, Xu:2017rfo, Yohana:2019ahg, Zhang:2019ipd, Tramonte:2020csa, Xu:2020uws}).

The first detection of the 21 cm signal in the IM regime was achieved by Chang et al. in 2010 \citep{Chang:2010jp}. Using the Green Bank Telescope (GBT), they detected a 3D 21 cm intensity field in the redshift range from $z=0.53$ to 1.12, which overlaps with $10,000$ galaxies in the DEEP2 galaxy survey \citep{Davis:2000vr}. This detection was the first verification that the 21 cm intensity field at $z\sim1$ traces the distribution of galaxies. After that, by expanding the GBT 21 cm IM survey in both sensitivity and spatial coverage, Masui et al. detected a cross-power spectrum at $z\sim0.8$ between 21 cm IM and galaxies in the WiggleZ Dark Energy Survey \citep{Masui:2012zc}, and Switzer et al. provided an upper limit on the 21 cm auto-power spectrum for the first time \citep{Switzer:2013ewa}. Lately, Anderson et al. reported a cross-power spectrum between the Parkes telescope 21 cm intensity maps and 2dF galaxy maps at $z\sim0.08$ \citep{Anderson:2017ert}. So far, no experiment has detected the 21 cm power spectrum in auto-correlation. Here, we consider six promising experiments in the redshift range $0<z<3$, namely, the Baryon acoustic oscillations In Neutral Gas Observations (BINGO) \cite{Battye:2012tg,Dickinson:2014wda}, the Five-hundred-meter Aperture Spherical radio Telescope (FAST) \cite{Nan:2011um,Smoot:2014oia}, the Square Kilometre Array Phase I (SKA1) \cite{Santos:2015gra,Braun:2015zta}, the Hydrogen Intensity and Real-time Analysis eXperiment (HIRAX) \cite{Newburgh:2016mwi}, the Canadian Hydrogen Intensity Mapping Experiment (CHIME) \cite{Newburgh:2014toa,Bandura:2014gwa}, and Tianlai \citep{2011SSPMA..41.1358C,2012IJMPS..12..256C}. This work aims to forecast what role the low-$z$ 21 cm IM experiments will play in constraining cosmological parameters. Note that the term ``low-$z$'' used here means that these 21 cm IM experiments only probe the late universe with $0<z<3$. For SKA1, it is known that it consists of the SKA1-Low array and the SKA1-MID array, and in this paper we only consider the 21 cm IM experiment performed by the SKA1-MID array (covering the redshift range $0.35<z<3$). Actually, the SKA1-Low array enables the 21 cm IM experiment to be extended to higher redshifts beyond $z=3$ \citep{Brax:2012cr,Hall:2012wd,Heneka:2018ins}.

The biggest challenge for all 21 cm IM experiments is the presence of foregrounds from the Milky Way and extragalactic point sources, which are orders of magnitude brighter than the redshifted 21 cm signal \citep{Alonso:2014sna}. Fortunately, the spectral structure of the foreground sources is basically smooth, so we can use some cleaning algorithms to suppress them to a level that the \textsc{H\,i} signal can be extracted in an unbiased way (e.g., Refs.~\citep{Wang:2005zj, Morales:2012kf, Parsons:2012qh, Liu:2014bba, Shaw:2014khi, Zhu:2016esh, Zuo:2018gzm, Carucci:2020enz, Cunnington:2020njn}). Even so, there will be some foreground contamination left. In this work, we will consider the residual foreground in the process of simulating the 21 cm IM data. We introduce a scaling factor $\varepsilon_{\rm FG}$ to characterize the foreground removal efficiency: $\varepsilon_{\rm FG}=1$ corresponds to no removal and $\varepsilon_{\rm FG}=0$ corresponds to perfect removal \citep{Bull:2014rha}. It is found that the constraint results are somewhat sensitive to the residual foreground contamination amplitude.

In this paper, we make a forecast for cosmological parameter estimation by using the simulated 21 cm IM data. We first compare the capabilities of constraining cosmological parameters for different 21 cm IM experiments. Then we combine the experiments with mainstream cosmological probes to investigate their help to break the degeneracies between the cosmological parameters. In addition, we investigate the effect of residual foreground on constraint results. Throughout this paper, we employ the Planck best-fit $\Lambda$CDM \citep{Aghanim:2018eyx} model for our fiducial cosmology, with $H_0=67.3\ \rm km\ s^{-1}\ Mpc^{-1}$, $\Omega_{\Lambda}=0.683$, $\Omega_{\rm m}=0.317$, $\Omega_{\rm b}=0.0495$, $\Omega_{\rm k}=0$, $\sigma_8=0.812$, and $n_{\rm s}=0.965$.

This paper is organized as follows. In Section \ref{sec2}, we give a detailed description of methodology. In Section \ref{sec2.1}, we introduce the Fisher forecast formalism for 21 cm IM. We further give a detailed description of the experimental configurations in Section \ref{sec2.2}, and describe data and method employed in Section \ref{sec2.3}. In Section \ref{sec3}, we show constraint results and discuss some relevant issues. Finally, we give our conclusions in Section \ref{sec4}.

\section{Methodology}\label{sec2}

\subsection{Fisher forecast formalism}\label{sec2.1}

Following the Fisher forecast formalism for 21 cm IM given in Ref.~\cite{Bull:2014rha}, we use the full 21 cm IM power spectrum to constrain the angular diameter distance, expansion rate, and structure growth rate, and then
use them to constrain cosmological parameters. Here, for completeness, we summarize the method and make some supplements.

In the IM regime, the spatial location of an observed pixel is given by 2D angular direction ${\boldsymbol{\theta}_p}$ and frequency ${\nu}_p$, i.e. \cite{Bull:2014rha}
\begin{align}
{\boldsymbol{r}}_{\perp} = r(z_{i})\big({\boldsymbol{\theta}}_{p}-{\boldsymbol{\theta}}_{i}\big);\ r_{\parallel} =r_{\nu}(z_{i})\big(\tilde{\nu}_{p}-\tilde{\nu}_{i}\big),
\end{align}
where the survey has been centered on $\big({\boldsymbol{\theta}}_{i},{\nu}_{i}\big)$, corresponding to a redshift bin centered at $z_{i}$. Here, $r(z)$ is comoving distance and $r_{\nu}(z)\equiv c(1+z)^2/H(z)$, where $c$ is the speed of light and $\tilde{\nu} \equiv \nu/\nu_{21}$ is the dimensionless frequency, with $\nu_{21}=1420.4\ \rm MHz$ being the rest-frame frequency of the 21 cm line. In the following, we will work in observational Fourier coordinates, $(\boldsymbol{q}={\boldsymbol{k}}_{\perp}r, y=k_{\parallel}r_{\nu})$, where $\boldsymbol{k}_{\perp}$ and $k_{\parallel}$ are the perpendicular and parallel components of the physical wave factor $\boldsymbol{k}$, respectively.

\subsubsection{Signal}

For a \textsc{H\,i} clump, the effective 21 cm brightness temperature can be written as \cite{Bull:2014rha}
\begin{align}
\overline{T}_{\rm b}(z)=\displaystyle{\frac{3hc^3A_{10}}{32\pi k_{\rm B}m_{\rm p}{\nu}_{21}^2}} \displaystyle{\frac{(1+z)^2}{H(z)}} \Omega_{\textsc{H\,i}}(z) \rho_{\rm c,0},
\end{align}
where $h$ is the Planck constant, $A_{10}=2.85\times10^{-15}\ {\rm s}^{-1}$ is the Einstein spontaneous emission coefficient \citep{Furlanetto:2006jb}, $k_{\rm B}$ is the Boltzmann constant, $m_{\rm p}$ is the proton mass, $\Omega_{\textsc{H\,i}}(z)$ is the fractional density of \textsc{H\,i}, and $\rho_{\rm c,0}=3H_{0}^{2}/8\pi G$ is the critical density of the universe today, with $H_{0}$ being the Hubble constant. We can write
\begin{align}
\Omega_{\textsc{H\,i}}(z)\equiv(1+z)^{-3}\rho_{\textsc{H\,i}}(z)/\rho_{\rm c,0},
\end{align}
where $\rho_{\textsc{H\,i}}(z)$ is the proper \textsc{H\,i} density, calculated by
\begin{align}
\rho_{\textsc{H\,i}}(z)=\int_{M_{\rm min}}^{M_{\rm max}}dM\displaystyle{\frac{dn}{dM}M_{\textsc{H\,i}}(M,z)},
\end{align}
where ${dn}/{dM}$ is the proper halo mass function and $M_{\textsc{H\,i}}(M,z)$ is the total \textsc{H\,i} mass in a halo of mass $M$. We take $M_{\textsc{H\,i}}(M,z)\propto M^{0.6}$, and the corresponding form of $\Omega_{\textsc{H\,i}}(z)$ is shown in Figure 20 in Ref.~\cite{Bull:2014rha}, which gives $\Omega_{\textsc{H\,i}}(z=0)=4.86\times10^{-4}$.

Considering the effect of RSDs \citep{Kaiser:1987qv}, the signal covariance can be written as \citep{Seo:2003pu, Bull:2014rha}
\begin{align}
\label{Noise cov}
C^{\rm S}(\boldsymbol{q},y)=\displaystyle{\frac{\overline{T}_{\rm b}^{2}(z_{i})\alpha_{\perp}^2\alpha_{\parallel}}{r^2r_{\nu}}}\left(b_{\textsc{H\,i}}+f\mu^2\right)^2\exp{\left(-k^2 \mu^2 \sigma_{\rm NL}^{2}\right)} \times P(k),
\end{align}
where $\alpha_{\perp}\equiv D_{\rm A}^{\rm fid}(z)/D_{\rm A}(z)$ and $\alpha_{\parallel}\equiv H(z)/H^{\rm fid}(z)$, with ``fid'' labeling the quantities calculated in the fiducial cosmology. $b_{\textsc{H\,i}}$ is the \textsc{H\,i} bias, given by
\begin{align}
b_{\textsc{H\,i}}(z)=\rho_{\textsc{H\,i}}^{-1} \int_{M_{\rm min}}^{M_{\rm max}}dM\displaystyle{\frac{dn}{dM}M_{\textsc{H\,i}}}\,b(M,z),
\end{align}
and the specific calculation can be found in Ref.~\cite{Xu:2014bya}. $f$ is the linear growth factor and $\mu\equiv k_{\parallel}/k$. For a dark energy model, the growth rate can be parameterized as $f(z)=\Omega_{\rm m}^{\gamma}(z)$, where $\Omega_{\rm m}(z)=\Omega_{\rm m}(1+z)^3H_0^2/H^2(z)$ and $\gamma\approx0.545$ for the $\Lambda$CDM model (within the framework of general relativity). In Eq.~(\ref{Noise cov}), the exponential term accounts for the ``Fingers of God'' effect due to velocity dispersion on small scales, and the non-linear dispersion scale is parameterized by $\sigma_{\rm NL}$. In our fiducial model, we take $\sigma_{\rm NL}=7\ \rm{Mpc}$ \citep{Li:2007rpa}, which corresponds to a non-linear scale of $k_{\rm NL}=0.14\ \rm{Mpc^{-1}}$. $P(k)=D^2(z)P(k,z=0)$, with $D(z)$ being the linear growth factor, which is related to $f(z)$ by \citep{Xu:2014bya}
\begin{align}
f=\displaystyle{\frac{d\ln{D(a)}}{d\ln{a}}}=-\displaystyle{\frac{1+z}{D(z)}}\displaystyle{\frac{dD(z)}{dz}},
\end{align}
and $P(k,z=0)$ being the matter power spectrum at $z=0$ that can be generated by {\tt CAMB} \citep{Lewis:1999bs}.

\subsubsection{Noise and effective beams}

We now turn to instrumental and sky noises. The noise covariance has the form \cite{Bull:2014rha}
\begin{align}
C^{\rm N}(\boldsymbol{q},y)=\displaystyle{\frac{\sigma_{\rm pix}^2V_{\rm pix}}{r^2r_{\nu}}} B_{\parallel}^{-1} B_{\perp}^{-2},
\end{align}
where $\sigma_{\rm pix}$ is the pixel noise temperature, $V_{\rm pix}=r^2{\rm FoV}\times r_{\nu}\delta\nu/\nu_{21}$ is the pixel volume, in which $\rm FoV$ is the field of view of each receiver and $\delta\nu$ is the bandwidth of
an individual frequency channel. The factors of $B_{\parallel}$ and $B_{\perp}$ describe the frequency and angular responses of the instrument, respectively.

For single-dish experiments (used as a collection of independent single dishes),
\begin{align}
\sigma_{\rm pix}=\displaystyle{\frac{T_{\rm sys}}{\sqrt{n_{\rm pol}t_{\rm tot}\delta\nu\left(\rm{FoV}/S_{\rm area}\right)}}}\displaystyle{\frac{\lambda^2}{A_{\rm e}\rm{FoV}}}\displaystyle{\frac{1}{\sqrt{N_{\rm d}N_{\rm b}}}},
\end{align}
and for interferometers,
\begin{align}
\sigma_{\rm pix}=\displaystyle{\frac{T_{\rm sys}}{\sqrt{n_{\rm pol}t_{\rm tot}\delta\nu\left(\rm{FoV}/S_{\rm area}\right)}}}
\displaystyle{\frac{\lambda^2}{A_{\rm e}\sqrt{\rm FoV}}}
\displaystyle{\frac{1}{\sqrt{n(\boldsymbol{u})N_{\rm b}}}},
\end{align}
where $T_{\rm sys}$ is the system temperature, $n_{\rm pol}=2$ is the number of polarization channels, $t_{\rm tot}$ is the total integration time, $S_{\rm area}$ is the survey area, $A_{\rm e}$ is the effective collecting area of each receiver, $N_{\rm d}$ is the number of dishes and $N_{\rm b}$ is the number of beams. For a dish reflector, $A_{\rm e}=\eta\pi(D_{\rm d}/2)^2$ and $\rm{FoV}\approx\theta_{\rm B}^2$, where $D_{\rm d}$ is the diameter of the dish, $\eta$ is the efficiency factor for which we adopt 0.7 in this work, and $\theta_{\rm B}\approx\lambda/D_{\rm d}$ is the full width at half-maximum (FWHM) of the beam. For a cylindrical reflector, $A_{\rm e}=\eta l_{\rm cyl}w_{\rm cyl}/N_{\rm feed}$ and ${\rm{FoV}}\approx 90^{\circ}\times\lambda/w_{\rm cyl}$, where $l_{\rm cyl}$ and $w_{\rm cyl}$ are the length and width of the cylinder, respectively, and $N_{\rm feed}$ is the number of feeds per cylinder. $n({\boldsymbol{u}})$ is the baseline density, calculated here from the actual array layout; note that the detailed calculation can be found in Ref.~\citep{Bull:2014rha}. Note also that, for dish interferometer, HIRAX, we neglect the baselines with $\left|u\right|\le1/\sqrt{\rm FoV}$, as these baselines are not independent \cite{Bull:2014rha}. For cylinder interferometers, CHIME and Tianlai, given the technical challenge and significant computational expense of correlating all baselines, we ignore the baselines shorter than the cylinder width. For the performance of Tianlai with no baseline cutting, we refer the reader to Refs.~\citep{Xu:2014bya,Zhang:2021yof}.

For BINGO, FAST, HIRAX, CHIME, and Tianlai, the system temperature is given by
\begin{align}
T_{\rm sys}=T_{\rm rec} + T_{\rm gal} + T_{\rm CMB},
\end{align}
where $T_{\rm rec}$ is the receiver noise temperature (the values for each experiment are listed in Table \ref{Telescope}), $T_{\rm gal}\approx25\ {\rm K}\times(408\ \rm{MHz}/\nu)^{2.75}$ is the contribution from the Milky Way, and $T_{\rm CMB}\approx2.73\ {\rm K}$ is the cosmic microwave background (CMB) temperature. For SKA1-MID, the system temperature is
\begin{align}
T_{\rm sys}=T_{\rm rec} + T_{\rm spl} + T_{\rm gal} + T_{\rm CMB},
\end{align}
where $T_{\rm spl}\approx3\ \rm K$ is the contribution from spill-over, and the receiver noise temperature is assumed to be \citep{Bacon:2018dui}
\begin{align}
\label{SKA1}
T_{\rm rec}=15\ {\rm K}+ 30\ {\rm K}\left(\nu/{\rm GHz}-0.75\right)^2.
\end{align}

In addition to noises, we need to consider the frequency and angular responses, which are capable of mixing frequency dependence and sky location. In the radial direction, the instrumental resolution is given by \citep{Battye:2012tg},
\begin{align}
B_{\parallel}(y)=\exp{\left(-\displaystyle{\frac{(y\delta\nu/\nu_{21})^2}{16\ln{2}}}\right)}.
\end{align}
Actually, due to the narrow channel bandwidths of modern radio receivers, the frequency resolution of IM surveys performs well. In transverse direction, the effective beam needs to be discussed separately. For single-dish experiments, it is calculated by
\begin{align}
B_{\perp}(\boldsymbol{q})=\exp{\left(-\displaystyle{\frac{(q\theta_{\rm B})^2}{16\ln{2}}}\right)},
\end{align}
and for interferometers, it has been accounted for by $n({\boldsymbol{u}})$.

\subsubsection{Residual foreground}

\begin{figure}[htbp]
\centering
\includegraphics[scale=0.5]{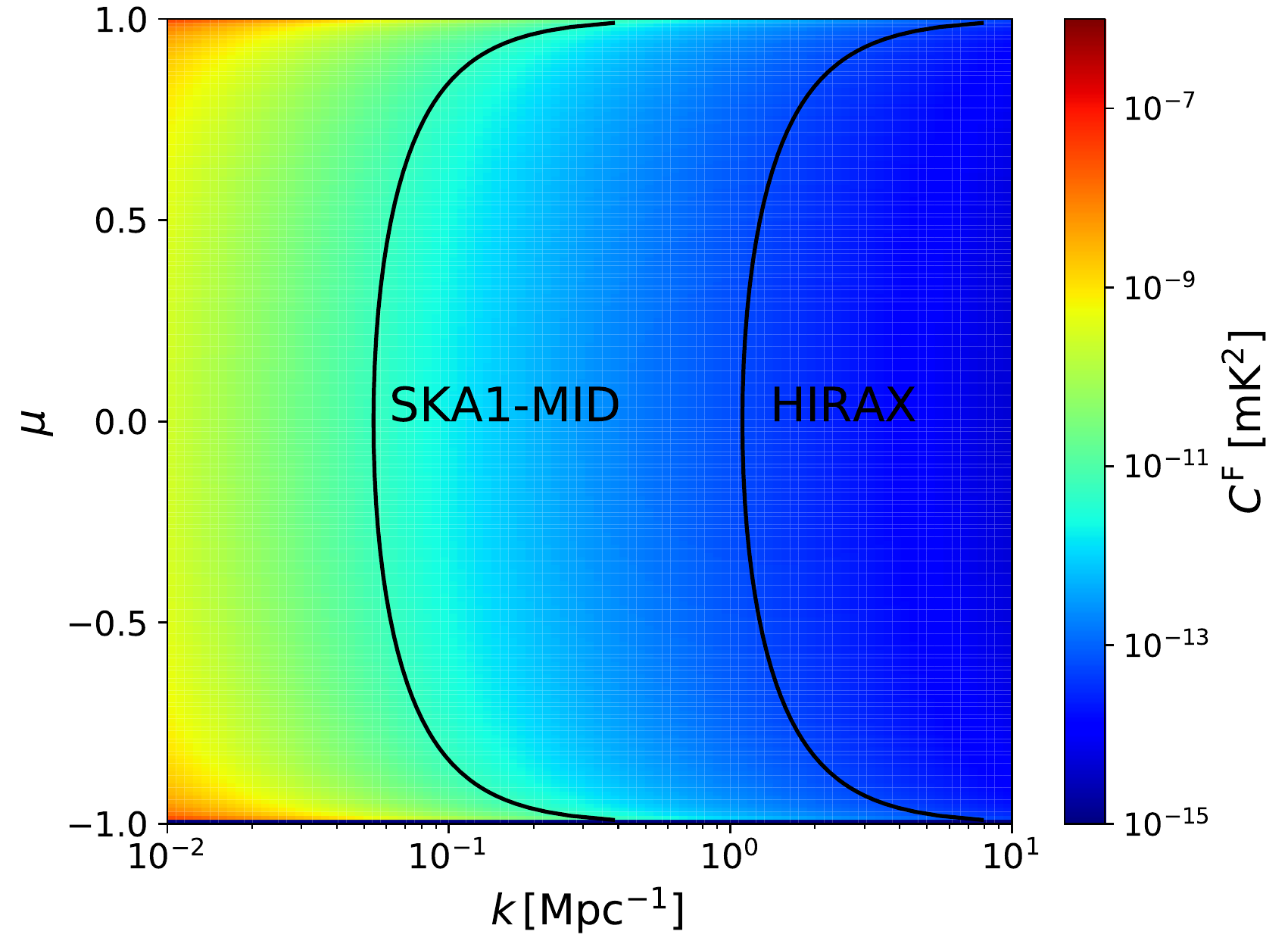}
\caption{The covariance of residual foreground at $z=1$ in the case of $\varepsilon_{\rm FG}=10^{-6}$, and the minimum angular scales that can be resolved by SKA1-MID and HIRAX.}
\label{fig:CF-FG6}
\end{figure}

Detection of the 21 cm signal is complicated by astrophysical foregrounds, primarily synchrotron from the Galaxy and unresolved radio point sources, which are orders of magnitude brighter than the signal. Fortunately, these foregrounds have a basically smooth spectral structure and hence we can use some cleaning algorithms to remove them (e.g., Refs. \citep{Wang:2005zj, Morales:2012kf, Parsons:2012qh, Liu:2014bba, Shaw:2014khi,Zhu:2016esh, Zuo:2018gzm, Carucci:2020enz, Cunnington:2020njn}). In this work, we assume that a cleaning algorithm has been applied and the covariance of residual foreground can be modeled as \citep{Santos:2004ju, Bull:2014rha}
\begin{align}
C^{\rm F}(\boldsymbol{q},y)=\varepsilon_{\rm FG}^2\sum_{X}A_{X}\left(\displaystyle{\frac{l_p}{2\pi q}}\right)^{n_{X}}\left(\displaystyle{\frac{\nu_p}{\nu_i}}\right)^{m_{X}},
\end{align}
where $\varepsilon_{\rm FG}$ is a scaling factor that parameterizes the efficiency of the foreground removal process: $\varepsilon_{\rm FG}=1$ corresponds to no foreground removal and $\varepsilon_{\rm FG}=0$ corresponds to perfect removal. We will need $\varepsilon_{\rm FG}\lesssim10^{-5}$ to extract the \textsc{H\,i} signal. Unless otherwise specified, we adopt a fiducial value of $\varepsilon_{\rm FG}=10^{-6}$. For a foreground $X$, $A_X$ is the amplitude, $n_X$ and $m_X$ are the angular scale and frequency power-law indexes, respectively. These parameters at $l_p=1000$ and $\nu_p=130\ \rm MHz$ are given in Table \ref{Tab:foreground}. We present the covariance of residual foreground at $z=1$ in Fig.~\ref{fig:CF-FG6}. Also shown in this figure are the minimum angular scales that can be resolved by SKA1-MID and HIRAX in the \textsc{H\,i} IM survey. Here, SKA1-MID and HIRAX are the representatives of single-dish and interferometric IM experiments, respectively. It can be seen that the foreground mainly affects the 21 cm signal on large scales, so the SKA1-MID in single-dish mode is most affected when the foreground removal efficiency is relatively low.
\begin{table}
\renewcommand\arraystretch{1.15}
\caption{Fiducial foreground parameters at $l_p=1000$ and $\nu_p=130\ \rm MHz$ \cite{Santos:2004ju}.}
\label{Tab:foreground}
\centering
\small
\begin{tabular}{p{4.3cm}|p{1.3cm}<{\centering} p{1cm}<{\centering}  p{1cm}<{\centering}}
\bottomrule[1pt]
Foreground                        & $A_X\ [\rm{mK^2}]$     & $n_X$      & $m_X$      \\
\hline
Extragalactic point sources       & 57.0                   & 1.1        & 2.07       \\
Extragalactic free-free           & 0.014                  & 1.0        & 2.10       \\
Galactic synchrotron              & 700                    & 2.4        & 2.80       \\
Galactic free-free                & 0.088                  & 3.0        & 2.15       \\
\bottomrule[1pt]
\end{tabular}
\end{table}

In addition, it is rather difficult to separate the smooth variation of the foreground in frequency from cosmological modes. Due to the frequency-dependence of the foreground subtraction, a minimum wave number cutoff in parallel direction, $k_{\rm FG}=2\pi/(r_{\nu}\Delta\tilde{\nu}_{\rm tot})$, is introduced, where $\Delta\tilde{\nu}_{\rm tot}$ is the total bandwidth of a survey~\cite{Bull:2014rha}, below which (namely, on larger scales in parallel direction) the cosmological information cannot be extracted.

\subsubsection{The Fisher matrix}

We define the total covariance as $C^{\rm T}=C^{\rm S}+C^{\rm N}+C^{\rm F}$, then the Fisher matrix for a cosmological parameter set $\{p\}$ is given by \cite{Bull:2014rha}
\begin{align}
\label{Fisher}
F_{ij}=\displaystyle{\frac{1}{2}}U_{\rm bin}\int\displaystyle{\frac{d^2qdy}{(2\pi)^3}}\ \displaystyle{\frac{\partial\ln{C^{\rm T}}}{\partial p_{i}}} \displaystyle{\frac{\partial\ln{C^{\rm T}}}{\partial p_{j}}},
\end{align}
where $U_{\rm bin}=S_{\rm area}\Delta \tilde{\nu}$, with $\Delta \tilde{\nu}$ being the dimensionless bandwidth in the redshift bin. We can rewrite Eq.~(\ref{Fisher}) in terms of wave vector
\begin{align}
F_{ij}=\displaystyle{\frac{1}{8\pi^2}}V_{\rm bin}\int_{-1}^{1}d\mu\int_{k_{\rm min}}^{k_{\rm max}}k^2dk \displaystyle{\frac{\partial\ln{C^{\rm T}}}{\partial p_{i}}} \displaystyle{\frac{\partial\ln{C^{\rm T}}}{\partial p_{j}}},
\end{align}
where $V_{\rm bin}=U_{\rm bin}r^2r_{\nu}$ is the physical volume. We assume that $b_{\textsc{H\,i}}$ is only redshift dependent, which is appropriate for large (linear) scales, so we impose a non-linear cut-off at $k_{\rm max}=k_{\rm NL}(1+z)^{2/(2+n_{\rm s})}$ \citep{Smith:2002dz}. In addition, the largest scale the survey can probe corresponds to a wave vector $k_{\rm min}=2\pi V_{\rm bin}^{-1/3}$. In this work, we choose the parameter set $\{p\}$ as $\{D_{\rm A}(z),H(z),[f\sigma_8](z), [b_{\textsc{H\,i}}\sigma_8](z), \sigma_{\rm NL}\}$. Note that we marginalize $[b_{\textsc{H\,i}}\sigma_8](z)$ and $\sigma_{\rm NL}$, and only use the angular diameter distance $D_{\rm A}(z)$, Hubble parameter $H(z)$, and RSD observable $[f\sigma_8](z)$, to constrain dark energy models.

\subsection{Experimental configurations}\label{sec2.2}
In this paper, we consider six experiments which are potentially suitable for 21 cm IM surveys, including BINGO, FAST, SKA1-MID, HIRAX, CHIME, and Tianlai. In this subsection, we will give a brief introduction to them.

$\textbf{BINGO:}$ The BINGO project\footnote{\url{http://www.bingotelescope.org}}, to be built in eastern Brazil, is a special-purpose radio telescope. It aims to detect the BAO in \textsc{H\,i} power spectrum, in the redshift interval $0.13<z<0.45$ \citep{Battye:2012tg,Dickinson:2014wda,Wuensche:2019znm,Wuensche:2021ebn}. The latest design of BINGO is a dual-mirror compact antenna telescope with a 40 m primary mirror and an offset focus, which has a receiver array containing 50-60 feed horns, with a focal length of 90 m.

$\textbf{FAST:}$ FAST\footnote{\url{https://fast.bao.ac.cn}} is a multi-beam single dish telescope with an aperture diameter of 500 m, built in Guizhou Province, Southwest China. FAST is believed to be the most sensitive single dish telescope currently in existence. It uses an active surface that adjusts shape to create parabolas in different directions, with an effective illuminating diameter of 300 m. It will be capable of covering the sky within a 40-degree angle from the zenith. The L-band receiver is designed with 19 beams, which will increase the survey speed \citep{Nan:2011um,Smoot:2014oia}.

$\textbf{SKA1-MID:}$ The SKA project\footnote{\url{https://www.skatelescope.org}} plans two stages of development: SKA1 is under construction, and SKA2 is planned to follow. In this paper, we only consider the SKA1-MID array, which will be built in the Northern Cape Province of South Africa as a mixed array of 133 15 m diameter dishes and 64 13.5 m dishes from the MeerKAT array \citep{Santos:2015gra,Braun:2015zta,Bacon:2018dui}. For simplicity, we approximate it to a 197 15 m dishes array. In addition, we only consider the {\it Wide Band 1 Survey} of SKA1-MID and use its single-dish (auto-correlation) mode. Note that the SKA1-MID array needs to balance multiple applications and realize high angular resolution, and so it is designed to have many long baselines and few short baselines. The design with many long baselines in interferometric mode can well realize the high angular resolution, which is suited to observe point sources but is not helpful in measuring the large-scale structure in the IM survey. To map the scales of interest with sufficient signal-to-noise in the IM survey, only the low angular resolution is needed, and the SKA1-MID array in interferometric mode does not have enough short baselines to perform this task \cite{Ansari:2018ury}. Therefore, the best way of measuring the large-scale structure in the \textsc{H\,i} IM survey is to use the SKA1-MID array in single-dish mode. Since the SKA1-MID array has a large number of dishes, it can guarantee a high survey speed for probing the \textsc{H\,i} signal and has the potential to probe cosmology over a wide range of scales with high signal-to-noise. At the same time, the interferometric data can be used to create high-resolution sky images that can be used for flux calibration in the IM survey as well as for other sciences.

$\textbf{HIRAX:}$ HIRAX\footnote{\url{https://hirax.ukzn.ac.za}}, a dish interferometer under construction in South Africa, will map nearly all of the southern sky in \textsc{H\,i} line emission over a frequency range of 400 to 800 MHz. It will be comprised of 1024 6 m diameter dishes, deployed in a $32\times32$ grid (7 m spacing) with the square sides aligned on the celestial cardinal directions \citep{Newburgh:2016mwi}. HIRAX is highly complementary to CHIME and will share many back-end technologies. Moreover, the Stage II 21 cm experiment suggested by Ref. \cite{Ansari:2018ury} is close to be a scaled version of HIRAX, consisting of a square array of $256\times256$ 6 m dishes, observing half the sky in the redshift range $2<z<6$. In this paper, we will only discuss HIRAX.

$\textbf{CHIME:}$ CHIME\footnote{\url{https://chime-experiment.ca}} is a close-packed cylinder interferometer being built in British Columbia, Canada, which aims to measure the BAO scale \citep{Newburgh:2014toa,Bandura:2014gwa}. It will consist of five adjacent cylindrical radio antennas with no moving parts, observing the sky which passes above it as the Earth rotates. Each cylinder is 20 m across and 100 m long ($20\times80\ \rm m^2 $ illuminated), with 256 dual-polarization feeds. It operates from 400-800 MHz, equivalent to mapping LSS between redshift $z = 0.77$ to 2.55. CHIME currently has four cylinders and has achieved another goal of detecting fast radio bursts \citep{Amiri:2019qbv,Amiri:2019bjk,CHIMEFRB:2020abu,CHIMEFRB:2021srp}. Note that we will only discuss the full-scale CHIME.

$\textbf{Tianlai:}$ The Tianlai project\footnote{\url{http://tianlai.bao.ac.cn}}, located in Hongliuxia, Balikun County, Xinjiang Autonomous Region, China, is a 21 cm IM experiment dedicated to the observation of LSS and the measurement of cosmological parameters such as the equation of state (EoS) of dark energy. It will consist of eight adjacent cylinders, each 15 m wide and 120 m long, with 256 dual-polarization feeds \citep{2011SSPMA..41.1358C,2012IJMPS..12..256C,Xu:2014bya}. It is designed to cover the redshift range of $0<z<2.55$, but we only consider $0.49<z<2.55$ due to the baseline cutting. Tianlai is now running in a pathfinder stage \citep{Li:2020ast}. Here, we will only discuss the full-scale Tianlai to be built in the future. Incidentally, the word \emph{Tianlai} means ``heavenly sound'' in Chinese.

The configuration parameters for BINGO, FAST, SKA1-MID, HIRAX, CHIME and Tianlai used in this paper are listed in Table \ref{Telescope}.

\begin{table}
\renewcommand\arraystretch{1.2}
\caption{Experimental configurations for BINGO, FAST, SKA1-MID, HIRAX, CHIME, and Tianlai.}
\label{Telescope}
\footnotesize
\centering
\begin{tabular}{p{1.5cm}|p{1.7cm}<{\centering} p{1.7cm}<{\centering} p{1.7cm}<{\centering} p{1.7cm}<{\centering} p{1.7cm}<{\centering} p{1.7cm}<{\centering}}
\bottomrule[1pt]
                                  & BINGO   & FAST     & SKA1-MID               & HIRAX    & CHIME    & Tianlai      \\
\hline
$z_{\rm min}$                     & 0.13    & 0        & 0.35               & 0.77     & 0.77      & 0.49        \\
$z_{\rm max}$                     & 0.45    & 0.35     & 3                  & 2.55     & 2.55      & 2.55        \\
$N_{\rm d}$                       & 1       & 1        & 197                & 1024     & N/A       & N/A          \\
$N_{\rm b}$                       & 50      & 19       & 1                  & 1        & 1         & 1           \\
$D_{\rm d}~[\rm m]$               & 40      & 300      & 15                 & 6        & N/A       & N/A          \\
$S_{\rm area}~[\rm{deg^2}]$        & 3000    & 20000    & 20000             & 15000    & 20000     & 20000       \\
$t_{\rm tot}~[\rm h]$             & 10000   & 10000    & 10000              & 10000    & 10000     & 10000       \\
$T_{\rm rec}~[\rm K]$             & 50      & 20       & Eq.~(\ref{SKA1})   & 50       & 50        & 50          \\
\bottomrule[1pt]
\end{tabular}
\end{table}

\subsection{Data and method}\label{sec2.3}
\begin{figure*}
\includegraphics[scale=0.32]{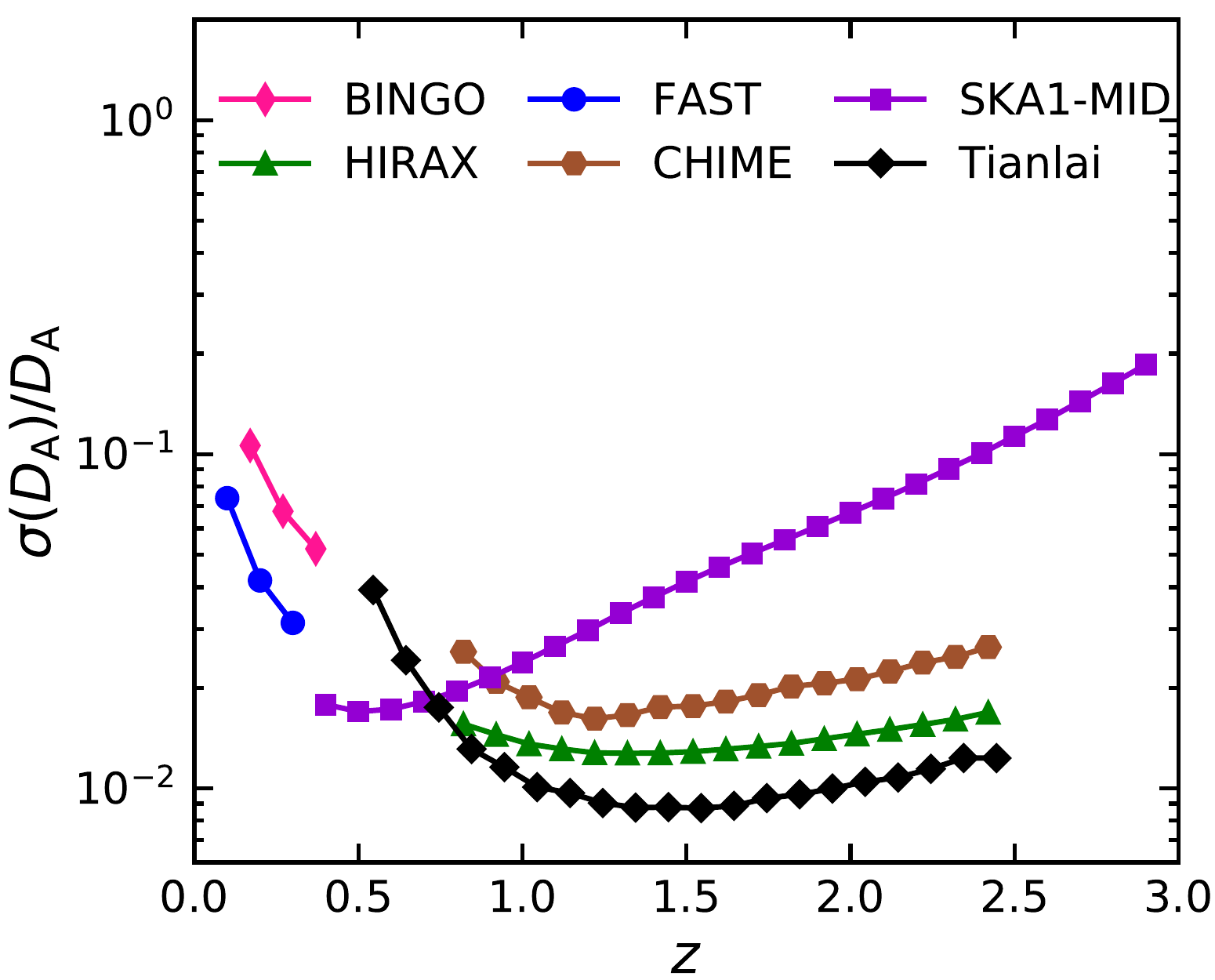}
\includegraphics[scale=0.32]{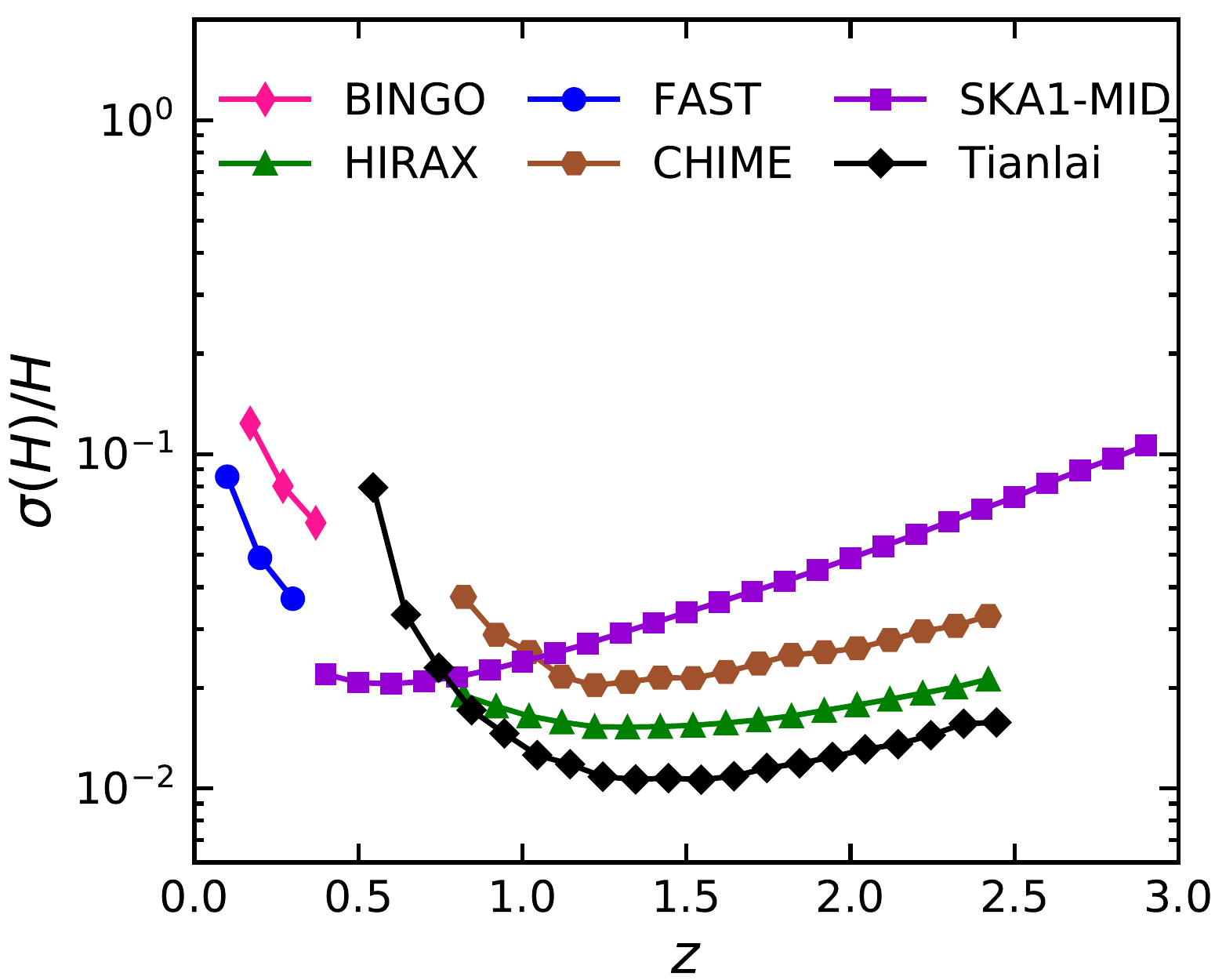}
\includegraphics[scale=0.32]{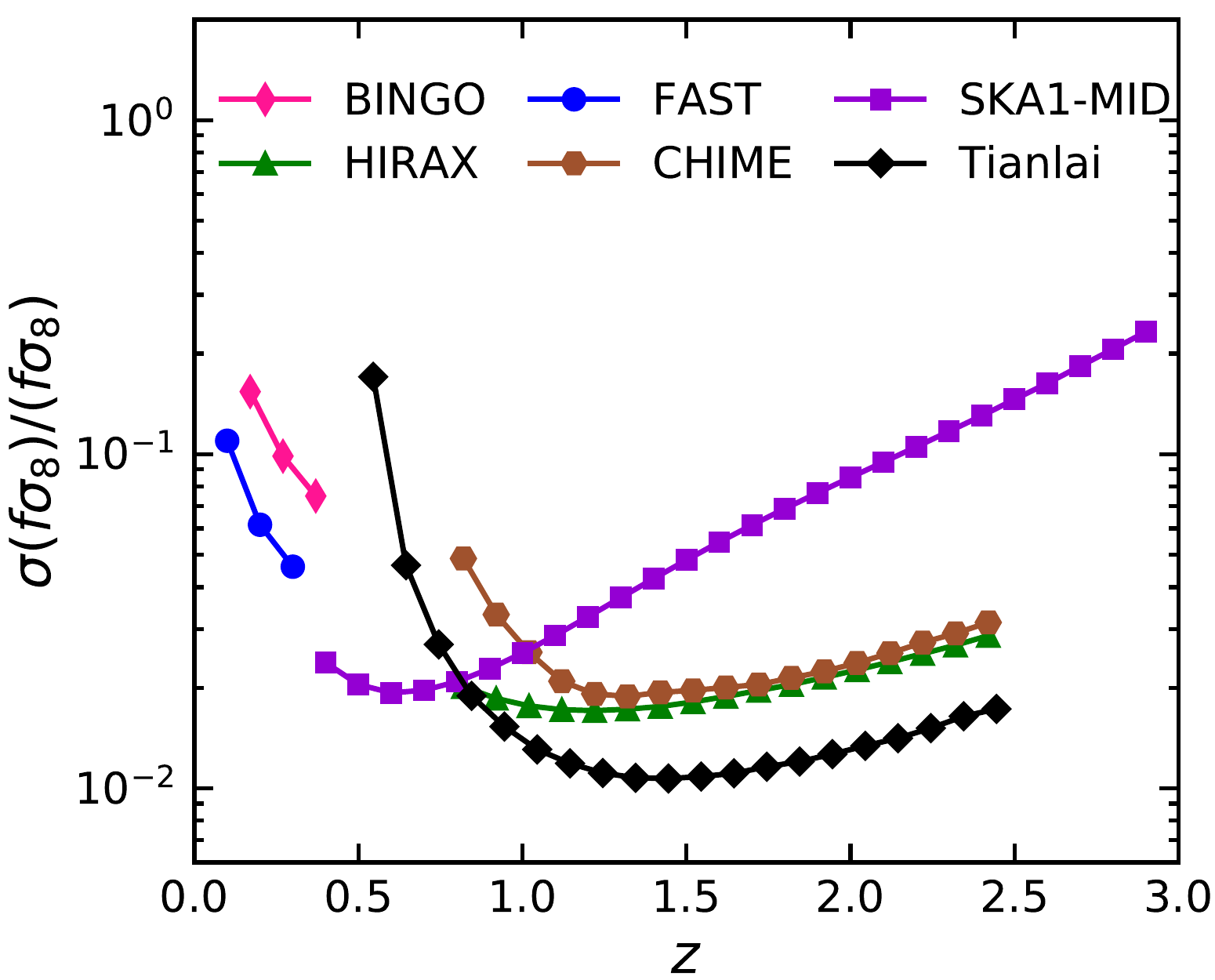}
\caption{Relative errors on $D_{\rm A}(z)$, $H(z)$, and $[f\sigma_8](z)$ in the case of $\varepsilon_{\rm FG}=10^{-6}$, as a function of redshift.}
\centering
\label{fig:frac err 6}
\end{figure*}

By performing measurements of the full 21 cm IM power spectrum, we can obtain the Fisher matrix for $\{D_{\rm A}(z),H(z),[f\sigma_8](z)\}$ in each redshift bin ($\Delta z=0.1$). When inverting the Fisher matrix, we can get the
covariance matrix that provides us with the forecast constraint on the chosen parameter set. Fig.~\ref{fig:frac err 6} shows the relative errors on $D_{\rm A}(z)$, $H(z)$ and $[f\sigma_8](z)$ for all experiments in the case of $\varepsilon_{\rm FG}=10^{-6}$. We see that for Tianlai, the errors grow rapidly at low redshifts, which is why we only consider $0.49<z<2.5$.

We use the Fisher matrices to constrain cosmological parameters by performing a Markov Chain Monte Carlo (MCMC) analysis \citep{Witzemann:2017lhi}. The cosmological parameters we sample include $H_0$, $\Omega_{\rm c}$, $\Omega_{\rm b}$, $\sigma_8$, $w$, $w_0$, and $w_a$, and we take flat priors for these parameters with ranges of $H_0\in[30, 100]\ \rm km\ s^{-1}\ Mpc^{-1}$, $\Omega_{\rm c}\in[0,1]$, $\Omega_{\rm b}\in[0,1]$, $\sigma_8\in[0,2]$, $w\in[-5,3]$, $w_0\in[-5,3]$, and $w_a\in[-5, 5]$. In the MCMC analysis, we also employ three current mainstream cosmological probes, namely, CMB, BAO, and SN. For the CMB data, we use the angular power spectra of Planck 2018 TT,TE,EE+lowE \citep{Aghanim:2018eyx}. For the BAO data, we consider the measurements from galaxy redshift surveys, including SDSS-MGS \citep{Ross:2014qpa}, 6dFGS \citep{Beutler:2011hx}, and BOSS DR12 \citep{Alam:2016hwk}. For the SN data, we use the latest sample from the Pantheon compilation \citep{Scolnic:2017caz}.

For a dark energy, the EoS is defined as $w(z)=p_{\rm de}(z)/\rho_{\rm de}(z)$. In this work, we consider three most typical cosmological models: (\romannumeral1) $\Lambda$CDM model---the standard cosmological model with $w(z)=-1$; (\romannumeral2) $w$CDM model---the simplest dynamical dark energy model with a constant EoS $w(z)=w$; (\romannumeral3) CPL model---the parameterized dynamical dark energy model with $w(z)=w_0+w_a z/(1+z)$ \citep{Chevallier:2000qy,Linder:2002et}.

\section{Results and discussion}\label{sec3}
We first report the constraint results from the simulated 21 cm IM data alone, and then combine these data with the mainstream cosmological probes. In the following discussions, we use $\sigma(\xi)$ and $\varepsilon(\xi)=\sigma(\xi)/\xi$ to represent the absolute and relative errors of the parameter $\xi$, respectively.
\begin{table*}[htbp]
\caption{The 1$\sigma$ errors on the cosmological parameters in the $\Lambda$CDM, $w$CDM, and CPL models, by using the 21 cm IM, CMB, CMB+21 cm IM data. Here, $H_0$ is in units of $\rm km\ s^{-1}\ Mpc^{-1}$. Note that the CMB-alone data can only provide lower limits for $H_0$ in $w$CDM and CPL, and an upper limit for $w_a$ in CPL, so it is not applicable for these cases to show 1$\sigma$ errors.}
\label{tab:21cm-result}
\footnotesize
\setlength\tabcolsep{1pt}
\renewcommand{\arraystretch}{1.5}
\centering
\begin{tabular}{lcccccccccccc}
\toprule[1pt]
\multicolumn{1}{l}{} &&\multicolumn{2}{c}{$\Lambda$CDM}&& \multicolumn{3}{c}{$w$CDM}&& \multicolumn{4}{c}{CPL}
\\
\cline{3-4} \cline{6-8} \cline{10-13} % \cline{1-1}
Data& &$\Omega_{\rm m}/10^{-3}$ &$H_0/10^{-1}$  &  &$\Omega_{\rm m}/10^{-3}$ &$H_0/10^{-1}$  &$w/10^{-2}$ &  &$\Omega_{\rm m}/10^{-3}$ &$H_0/10^{-1}$ &$w_0/10^{-2}$ &$w_a/10^{-1}$
\\
\hline

BINGO
        &
        & $39$
        & $29$
        &
        & $48$
        & $29$
        & $13$
        &
        & $41$
        & $31$
        & $37$
        & $14$
        \\
FAST
        &
        & $25$
        & $17$
        &
        & $35$
        & $21$
        & $7.5$
        &
        & $36$
        & $22$
        & $23$
        & $9.5$
        \\
SKA1-MID
        &
        & $7.0$
        & $5.5$
        &
        & $6.8$
        & $6.6$
        & $3.2$
        &
        & $15$
        & $11$
        & $11$
        & $4.2$
        \\
CHIME
        &
        & $5.3$
        & $4.0$
        &
        & $6.4$
        & $8.1$
        & $4.6$
        &
        & $33$
        & $30$
        & $28$
        & $7.9$
        \\
HIRAX
        &
        & $4.6$
        & $3.2$
        &
        & $4.7$
        & $5.8$
        & $3.0$
        &
        & $22$
        & $18$
        & $18$
        & $5.4$
        \\
Tianlai
        &
        & $2.7$
        & $2.0$
        &
        & $3.2$
        & $4.3$
        & $2.4$
        &
        & $20$
        & $16$
        & $16$
        & $4.7$
        \\
CMB
        &
        & $8.3$
        & $5.9$
        &
        & $34$
        & N/A
        & $25$
        &
        & $41$
        & N/A
        & $48$
        & N/A
        \\
CMB+BINGO
        &
        & $8.1$
        & $5.8$
        &
        & $15$
        & $15$
        & $4.4$
        &
        & $16$
        & $15$
        & $25$
        & $10$
        \\
CMB+FAST
        &
        & $7.9$
        & $5.6$
        &
        & $12$
        & $11$
        & $2.9$
        &
        & $12$
        & $11$
        & $21$
        & $8.5$
        \\
CMB+SKA1-MID
        &
        & $5.2$
        & $3.7$
        &
        & $5.4$
        & $4.4$
        & $1.3$
        &
        & $8.7$
        & $7.4$
        & $8.0$
        & $2.5$
        \\

CMB+CHIME
        &
        & $4.3$
        & $3.1$
        &
        & $4.8$
        & $4.3$
        & $1.7$
        &
        & $16$
        & $14$
        & $14$
        & $4.2$
        \\
CMB+HIRAX
        &
        & $3.7$
        & $2.6$
        &
        & $4.0$
        & $3.5$
        & $1.4$
        &
        & $14$
        & $13$
        & $13$
        & $3.6$
        \\
CMB+Tianlai
        &
        & $2.4$
        & $1.7$
        &
        & $2.7$
        & $2.7$
        & $1.4$
        &
        & $13$
        & $11$
        & $11$
        & $3.1$
        \\
\bottomrule[1pt]
\end{tabular}
\end{table*}

\begin{figure*}
\includegraphics[width=6.3cm,height=6.3cm]{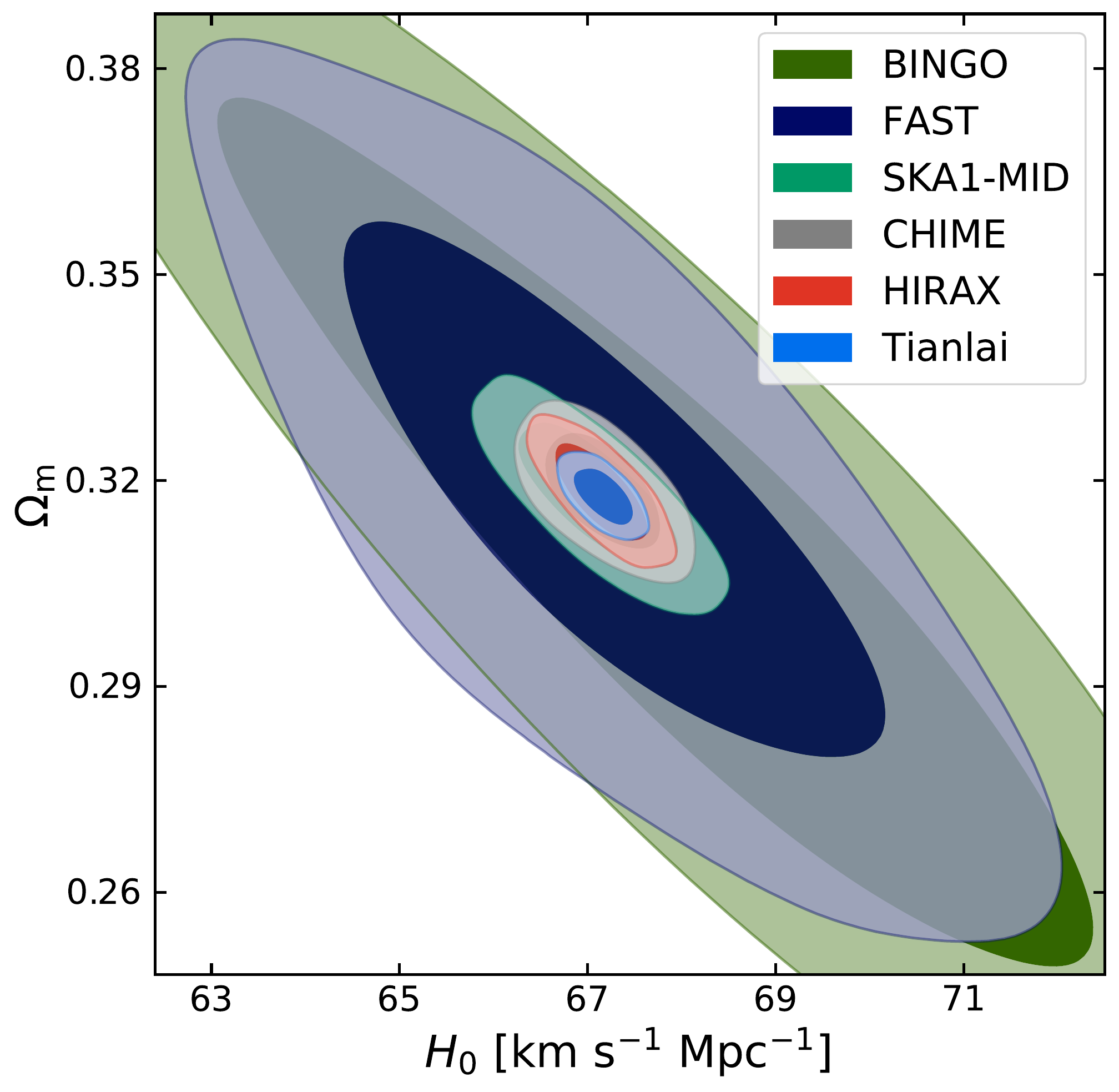}
\includegraphics[width=6.3cm,height=6.3cm]{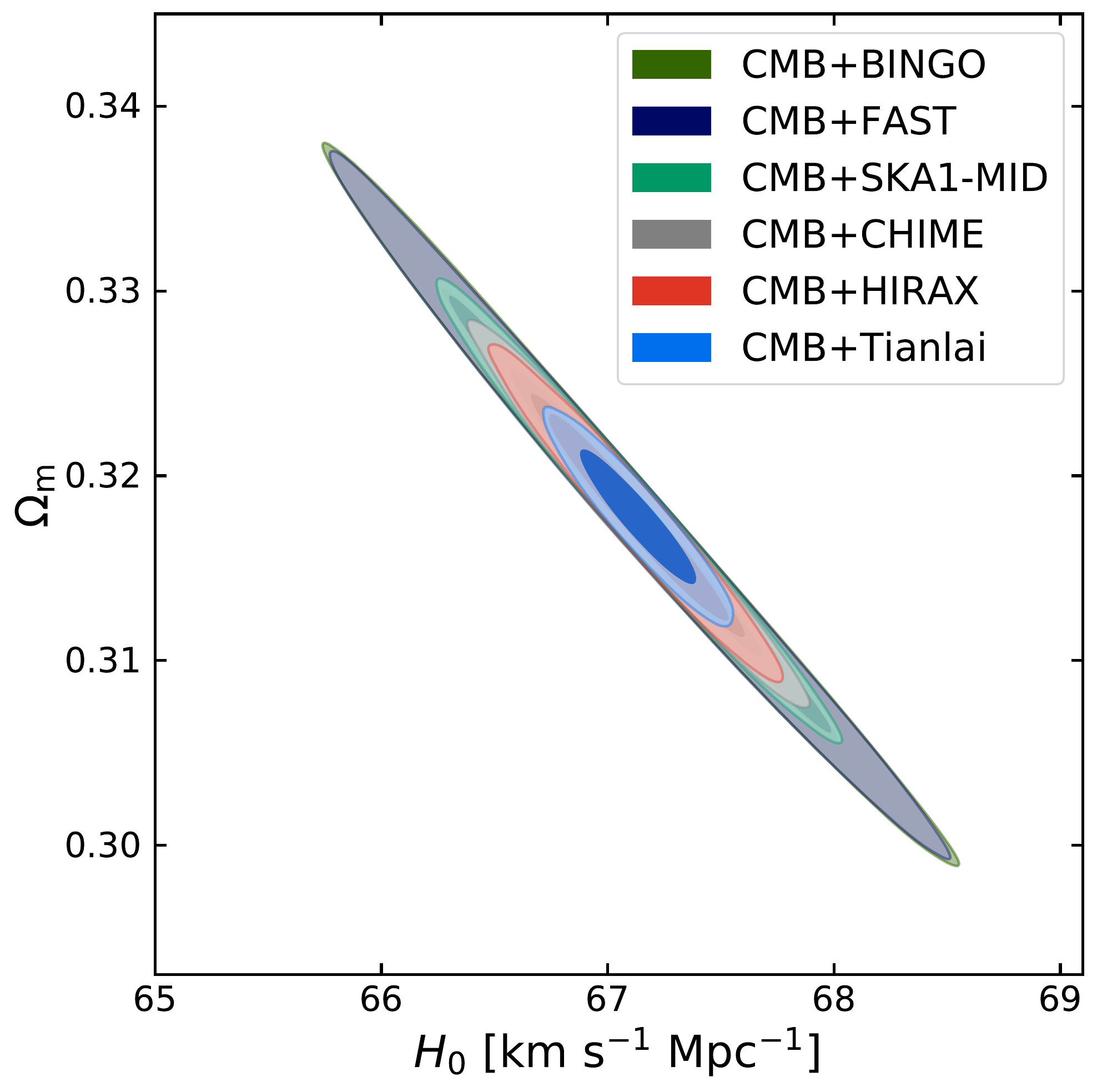}
\centering
\caption{Left: constraints ($68.3\%$ and $95.4\%$ confidence level) on the $\Lambda$CDM model by using BINGO, FAST, SKA1-MID, HIRAX, CHIME, and Tianlai. Right: constraints on the $\Lambda$CDM model by using CMB+BINGO, CMB+FAST, CMB+SKA1-MID, CMB+HIRAX, CMB+CHIME, and CMB+Tianlai.}
\label{fig:21cm-LCDM}
\end{figure*}

\begin{figure*}
\includegraphics[width=7.5cm,height=7.5cm]{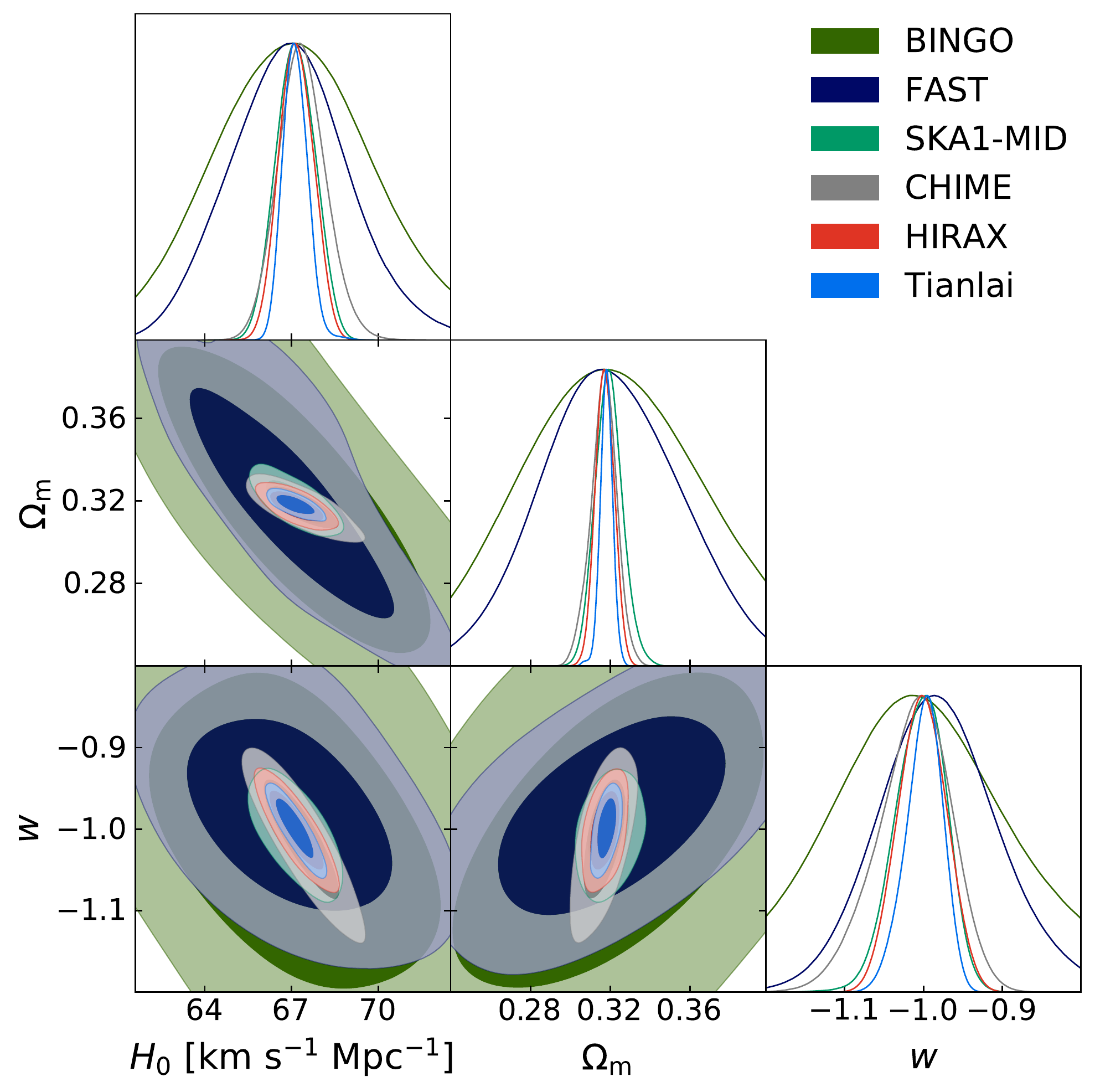}
\includegraphics[width=7.5cm,height=7.5cm]{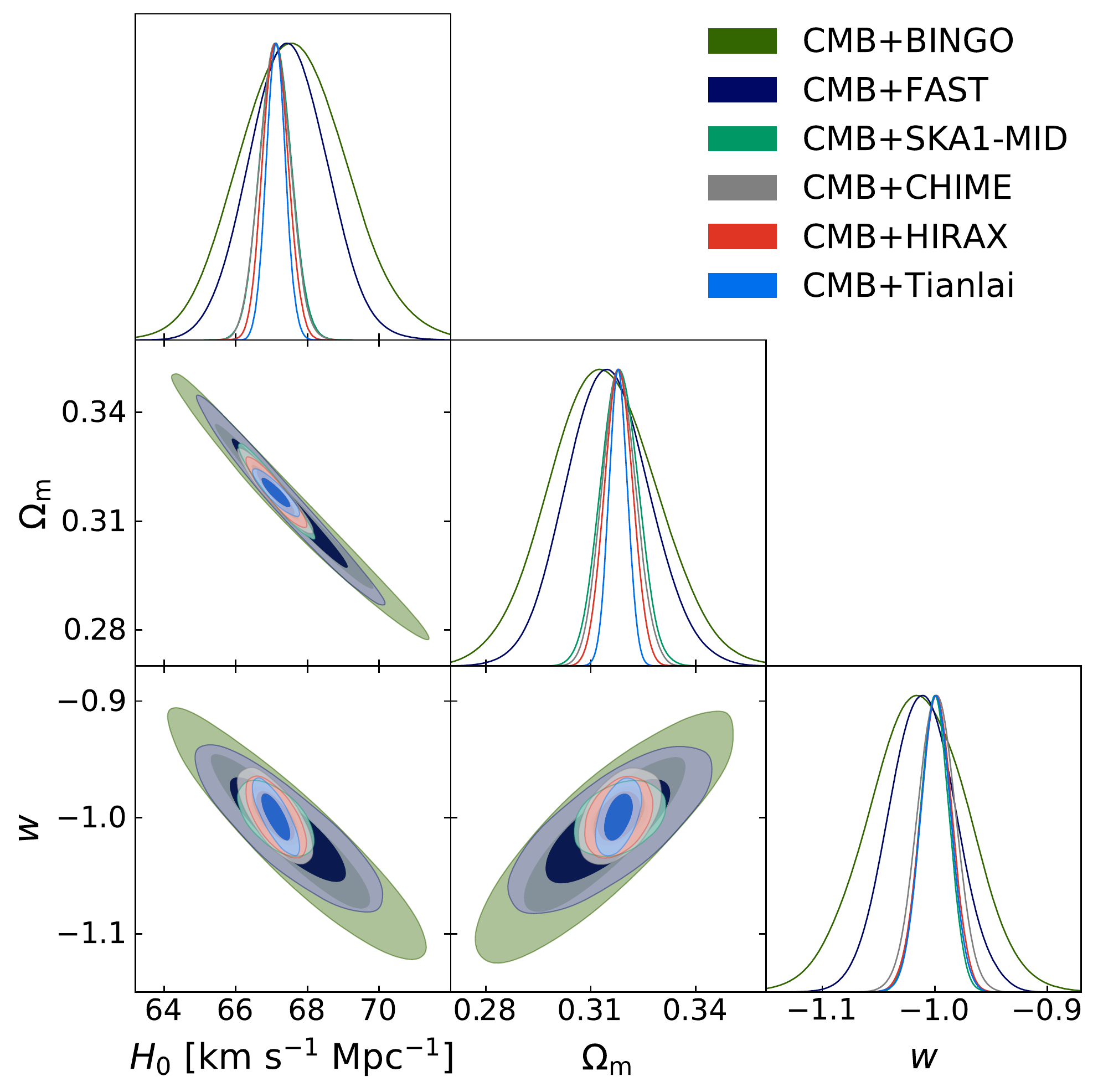}
\centering
\caption{Same as Fig.~\ref{fig:21cm-LCDM} but for the $w$CDM model.}
\label{fig:21cm-wCDM}
\end{figure*}

\begin{figure*}
\includegraphics[width=6.3cm,height=6.3cm]{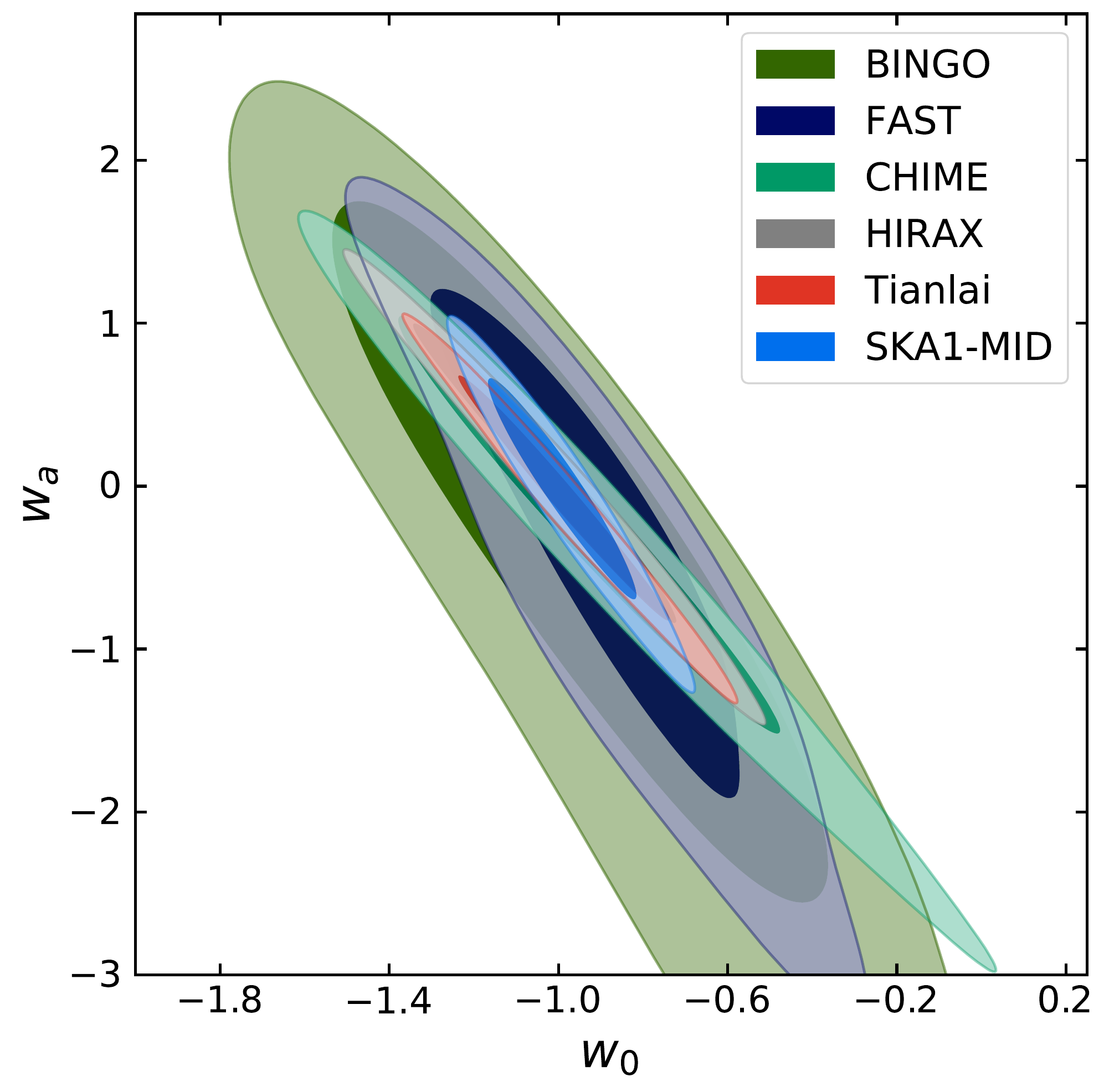}
\includegraphics[width=6.3cm,height=6.3cm]{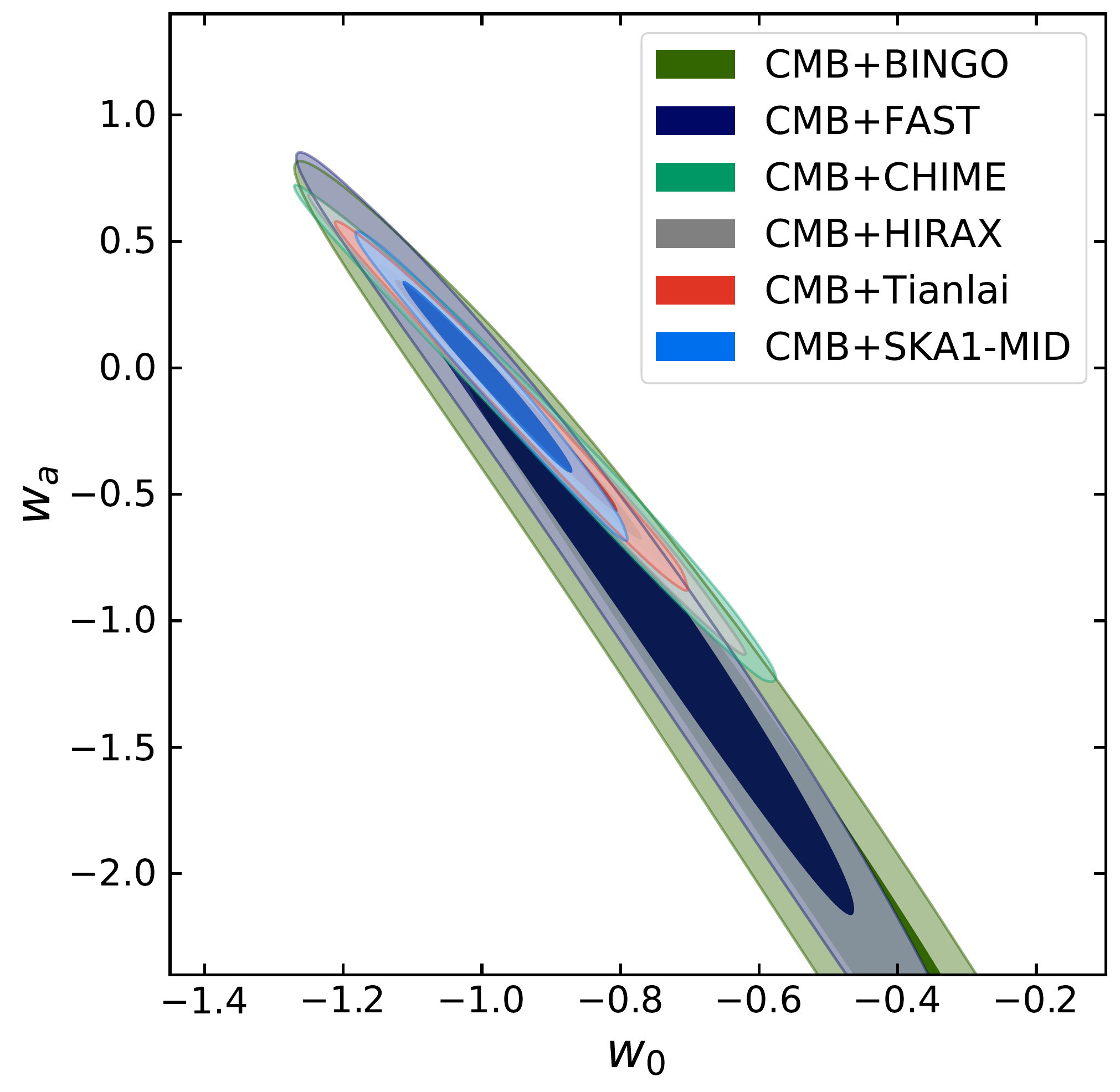}
\centering
\caption{Same as Fig.~\ref{fig:21cm-LCDM} but for the CPL model.}
\label{fig:21cm-CPL}
\end{figure*}

\subsection{21 cm IM experiments alone}\label{sec3.1}
In this subsection, we report the constraint results from the 21 cm IM experiments alone. The $1\sigma$ and $2\sigma$ posterior distribution contours for various model parameters are shown in the left panel of Figs.~\ref{fig:21cm-LCDM}--\ref{fig:21cm-CPL}, and the $1\sigma$ errors for the marginalized parameter constraints are summarized in Table~\ref{tab:21cm-result}. From the contours, BINGO and FAST are always weak in constraining cosmological parameters, because their redshift ranges are relatively narrow, with $0.13-0.45$ for BINGO and $0-0.35$ for FAST. Although the redshift range coverages are similar, FAST gives tighter constraints than BINGO, due to its larger survey area and aperture size, as well as lower receiver noise, which is consistent with the results given in Ref.~\cite{Zhang:2021yof}. In contrast, SKA1-MID, HIRAX, CHIME, and Tianlai can significantly tighten the constraints, mainly due to their wide redshift ranges and high spatial resolution. Also, the 21 cm IM surveys at redshifts $0.5<z<2.5$ are expected to provide a sensitive probe of cosmic dark energy, which is particularly true around the onset of acceleration at $z\approx1$ \citep{Chang:2007xk,Chang:2010jp}.

In the $\Lambda$CDM model, Tianlai gives the best constraints, $\sigma(\Omega_{\rm m})=0.0027$, $\sigma(H_0)=0.20\ \rm km\ s^{-1}\ Mpc^{-1}$, $\varepsilon(\Omega_{\rm m})=0.85\%$, and $\varepsilon(H_0)=0.30\%$, which means that Tianlai alone can achieve precision cosmology. The constraints of HIRAX, CHIME, and SKA1-MID on $\Omega_{\rm m}$ and ${H_0}$ decrease successively, but all of them are better than the Planck 2018 results \citep{Aghanim:2018eyx}. In the $w$CDM model, Tianlai still performs best, giving $\sigma(\Omega_{\rm m})=0.0032$, $\sigma(H_0)=0.43\ \rm km\ s^{-1}\ Mpc^{-1}$, and $\sigma(w)=0.024$. It is worth noting that SKA1-MID provides a tighter constraint on $w$ than CHIME, which seems to contradict the fact that CHIME is superior to SKA1-MID in $\Lambda$CDM. From Fig.~\ref{fig:frac err 6}, we can see that SKA1-MID performs very well at $0.35<z<0.77$, where dark energy dominates the evolution of the universe, so surveys in this range help to better constrain the dynamical dark energy EoS parameters. However, CHIME still has a better constraint on $\Omega_{\rm m}$ than SKA1-MID. This is because CHIME performs better than SKA1-MID at $1.0<z<2.5$, which is in the matter-dominated era of the universe. When the dark-energy EoS becomes evolutionary, SKA1-MID expands its advantage in constraining EoS parameters. In the CPL model, SKA1-MID offers the most precise constraints, $\sigma(\Omega_{\rm m})=0.015$, $\sigma(H_0)=1.1\ \rm km\ s^{-1}\ Mpc^{-1}$, $\sigma(w_0)=0.11$, and $\sigma(w_a)=0.42$.

The above analysis shows that the compact interferometers with high spatial resolution will have great advantages in constraining cosmological parameters, especially the Tianlai cylinder array. Note that for Tianlai, we have ignored the baselines shorter than 15\,m, and we only consider a redshift range of $0.49<z<2.55$. Nevertheless, Tianlai gives the best constraint precision in $\Lambda$CDM and $w$CDM. It is worth mentioning that the performance of FAST is undoubtedly very good at low redshifts, so it has the potential to combine with other 21 cm IM experiments, such as SKA1-MID. From the relative errors of cosmological observable in Fig. \ref{fig:frac err 6}, we propose a novel survey strategy, specifically, FAST ($0<z<0.35$) + SKA1-MID ($0.35<z<0.77$) + Tianlai ($0.77<z<2.55$). Here, Tianlai can be replaced by HIRAX or CHIME. This strategy gives full play to the advantages of relevant experiments, and we plan to investigate it in future works.

In the future, FAST will be expanded to cover a redshift range of $0.5<z<2.5$, and the corresponding survey strategies have been proposed \cite{Smoot:2014oia}. Moreover, the proposed Stage \uppercase\expandafter{\romannumeral2} 21 cm experiment \citep{Ansari:2018ury} will be able to survey larger volumes with higher resolution, potentially measuring the cosmological parameters to extremely high precision.

\subsection{Combination with CMB}\label{sec3.2}
In this subsection, we combine the 21 cm IM data with the current CMB data to study its help in improving the cosmological parameter constraints. We show the results in Table \ref{tab:21cm-result} and the right panel of Figs.~\ref{fig:21cm-LCDM}--\ref{fig:21cm-CPL}.

The CMB-alone constraints will lead to severe degeneracies between cosmological parameters in the extended models of  $\Lambda$CDM, as shown in the eighth row of Table~\ref{tab:21cm-result} and Fig.~\ref{fig:CBS}. However, the 21 cm IM data can effectively break the degeneracies. For example, in the $\Lambda$CDM model, CMB+Tianlai improves the constraints on $\Omega_{\rm m}$ and $H_0$ by $71\%$ and $71\%$, respectively, compared with CMB alone. In the $w$CDM model, CMB+SKA1-MID provides a slightly tighter constraint on $w$ than CMB+Tianlai, which shows that combining with CMB can highlight the advantage of SKA1-MID in constraining dynamical dark-energy EoS parameters. Importantly, CMB+SKA1-MID gives a very tight constraint on $w$, $\sigma(w)=0.013$, which is close to the level of precision cosmology. In the CPL model, CMB+SKA1-MID also gives the best constraints, $\sigma(w_0)=0.080$ and $\sigma(w_a)=0.25$, which are comparable to the CMB+BAO+SN results, as shown in the third row of Table \ref{tab:21cm-result}. Note that in Ref.~\cite{Xu:2014bya}, Xu et al. predicted that CMB+Tianlai can constrain $w_0$ and $w_a$ to $\sigma(w_0)=0.082$ and $\sigma(w_a)=0.21$, which are $(0.11-0.082)/0.11=25\%$ and $(0.31-0.21)/0.31=32\%$ better than our predictions, and are comparable to the results of CMB+SKA1-MID. This is mainly because they used all the baselines and assumed perfect foreground removal.

In conclusion, as late-time measurements, the 21 cm IM surveys are expected to break the parameter degeneracies inherent in the early-time CMB observations and thus achieve excellent constraint precision. Combining the 21 cm IM surveys with other late-time observations has been studied recently. In Ref.~\cite{Jin:2021pcv}, Jin et al. predicted that the synergy between the gravitational-wave standard siren observations based on Taiji + Cosmic Explorer and the 21 cm IM surveys based on SKA1-MID could achieve $\sigma(w_0)=0.077$ and $\sigma(w_a)=0.30$, which are $64\%$ and $81\%$ better than the results of Taiji + Cosmic Explorer. The above analysis shows that it is very beneficial to combine with 21 cm IM experiments for other observations in improving cosmological parameter estimation.

\subsection{Comparison with CMB+BAO+SN}\label{sec3.3}
\begin{figure*}[!htbp]
\includegraphics[width=6cm,height=5.3cm]{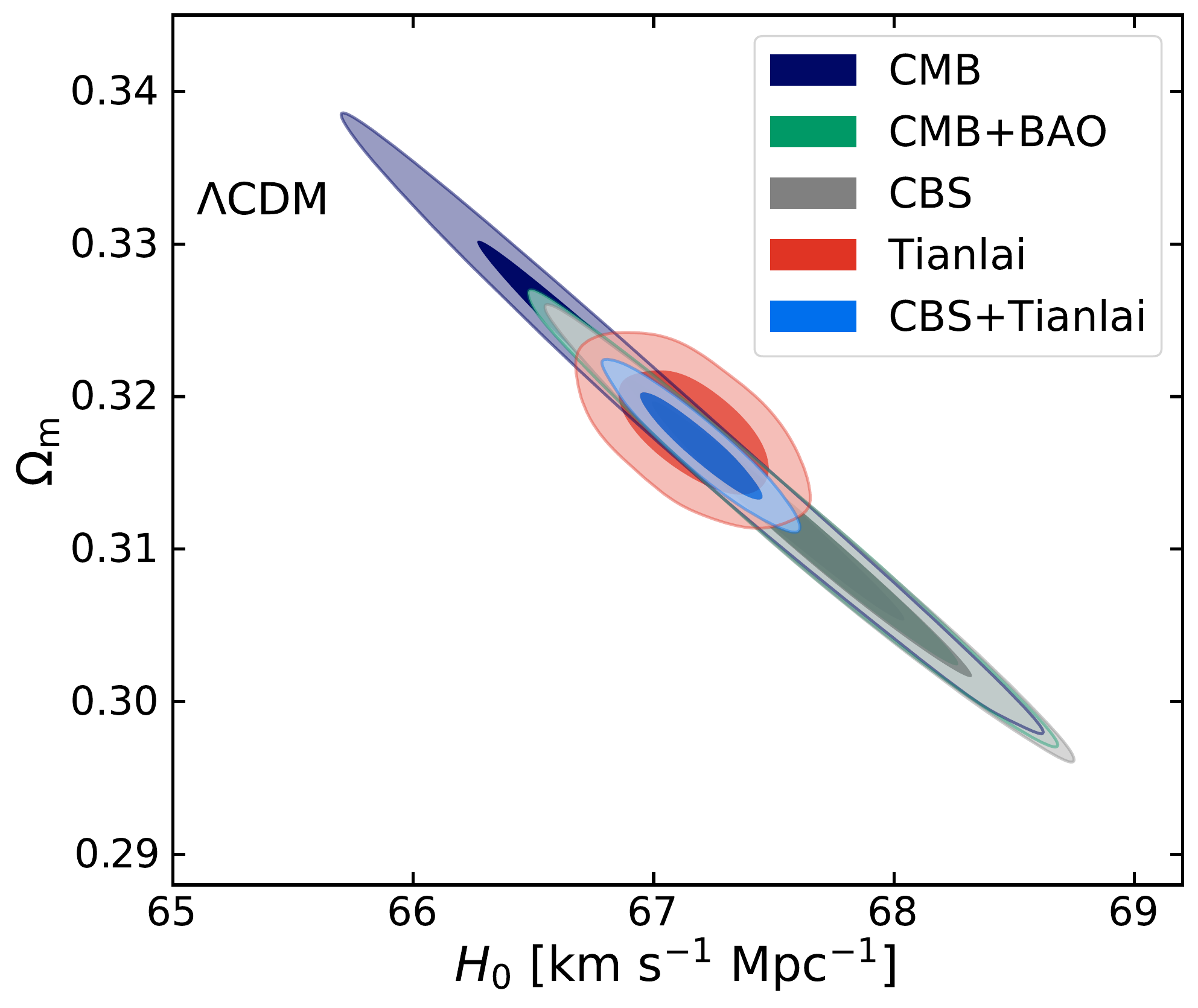}
\includegraphics[width=6cm,height=5.3cm]{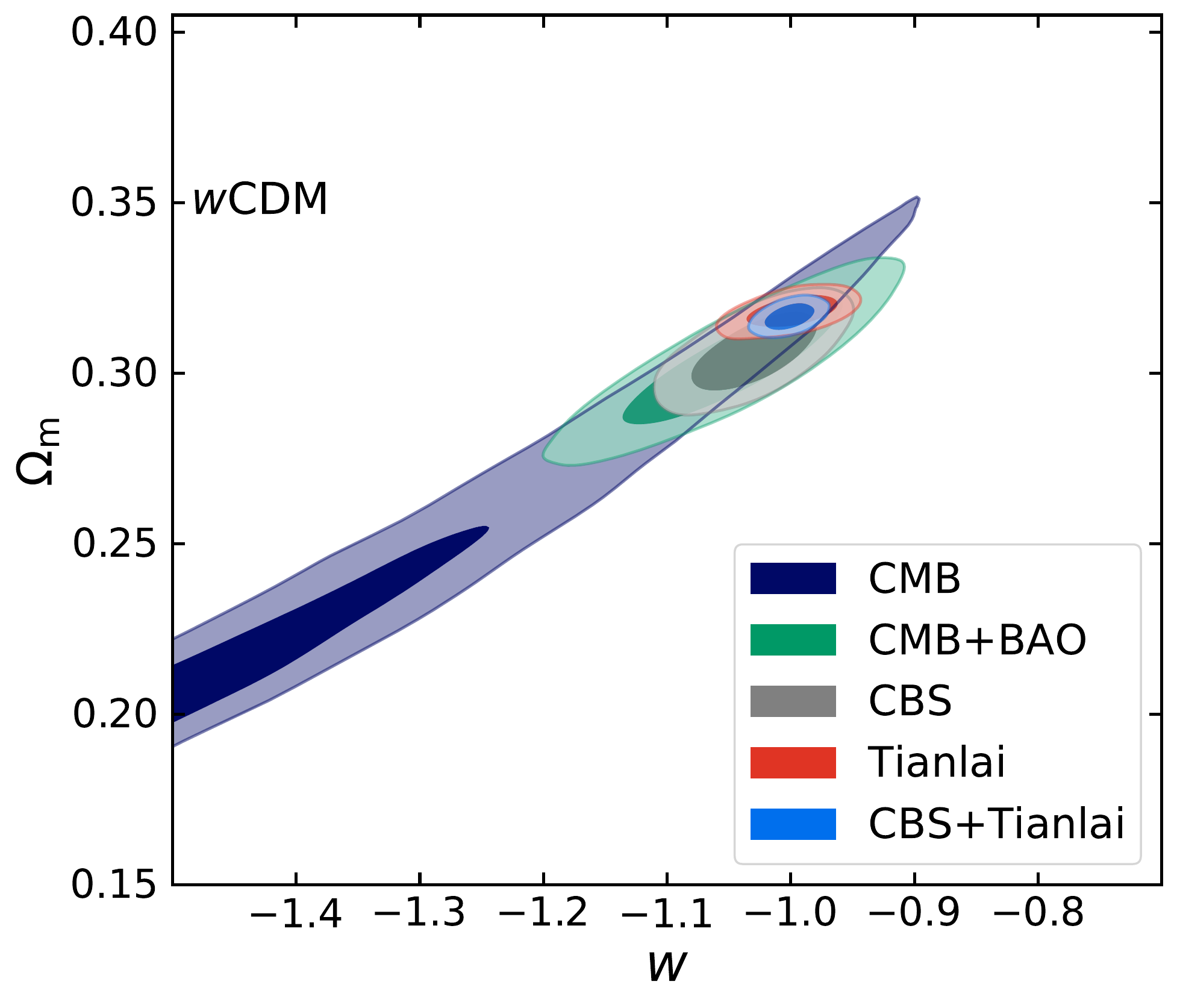}
\includegraphics[width=6cm,height=5.3cm]{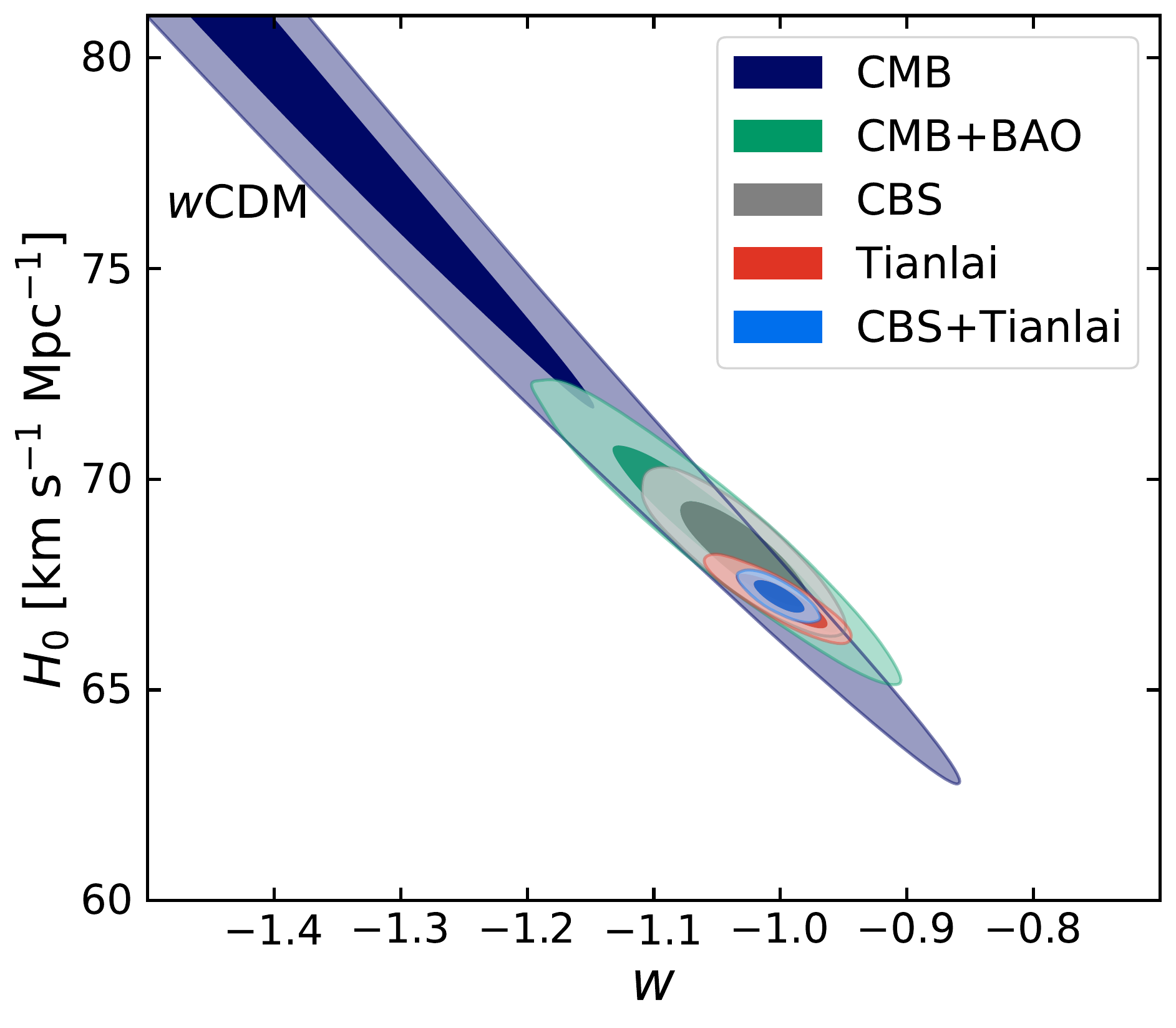}
\includegraphics[width=6cm,height=5.3cm]{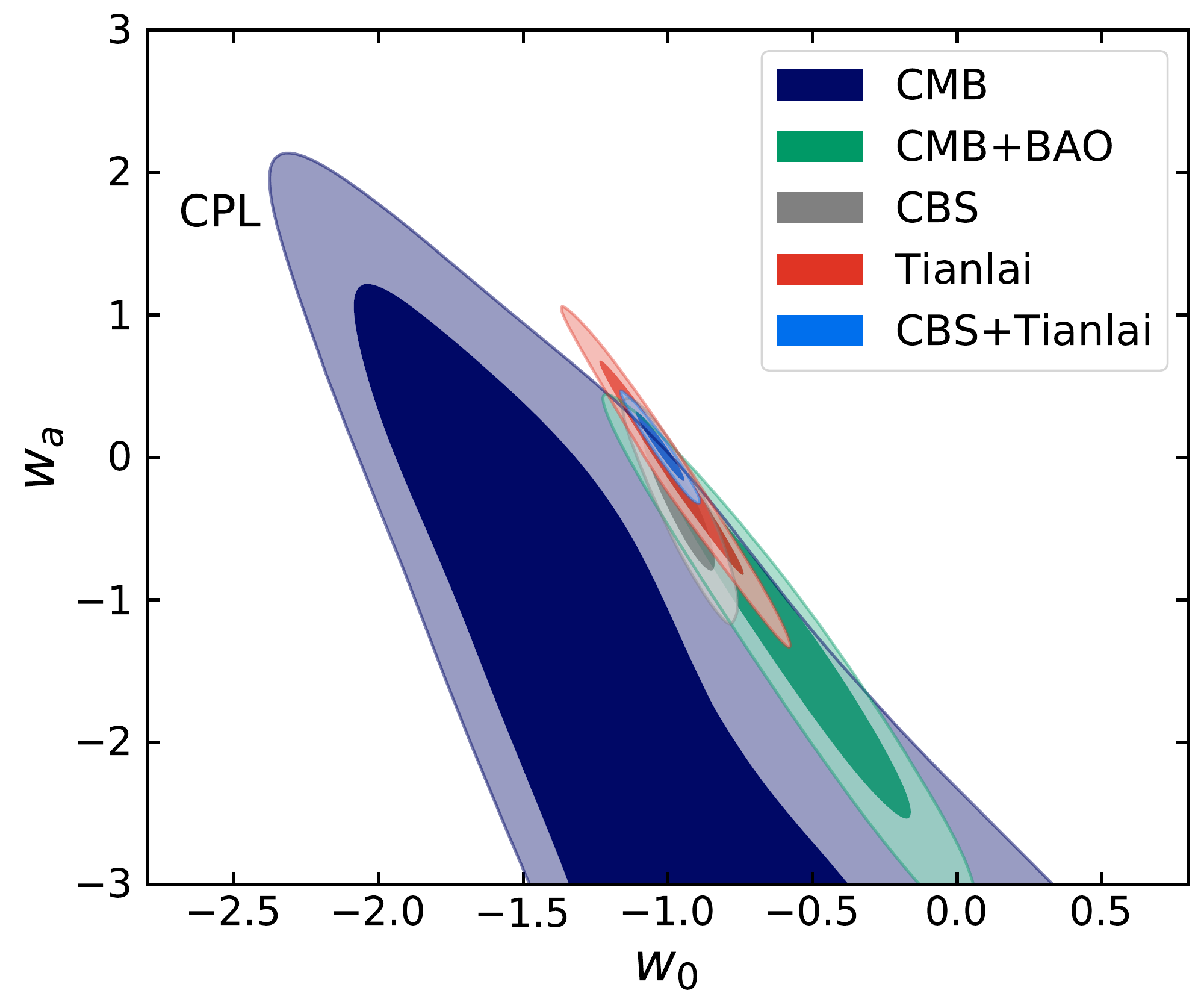}
\centering
\caption{Constraints (68.3\% and 95.4\% confidence level) on the $\Lambda$CDM, $w$CDM, and CPL models, by using CMB, CMB+BAO, CBS, Tianlai, and CBS+Tianlai. Here, CBS stands for CMB+BAO+SN.}
\label{fig:CBS}
\end{figure*}

In this subsection, we compare the constraints of the Tianlai-alone data with those of the CMB, CMB+BAO, and CMB+BAO+SN data, to illustrate the role of future 21 cm IM experiments in constraining cosmological parameters. We also investigate the constraints that the combination CMB+BAO+SN+Tinalai can provide. The results are shown in Fig.~\ref{fig:CBS} and Table~\ref{tab:CBS-result}. Note that for convenience, we use CBS to represent CMB+BAO+SN.

\begin{table*}[htbp]
\caption{The 1$\sigma$ errors on the cosmological parameters in the $\Lambda$CDM, $w$CDM, and CPL models, by using CMB, CMB+BAO, CBS, Tianlai, and CBS+Tianlai. Here, CBS stands for CMB+BAO+SN, and $H_0$ is in units of $\rm km\ s^{-1}\ Mpc^{-1}$. Note that the CMB-alone data can only provide lower limits for $H_0$ in $w$CDM and CPL, and an upper limit for $w_a$ in CPL, so it is not applicable for these cases to show 1$\sigma$ errors.}
\label{tab:CBS-result}
\footnotesize
\setlength\tabcolsep{1.5pt}
\renewcommand{\arraystretch}{1.5}
\centering
\begin{tabular}{lcccccccccccc}
\toprule[1pt]
\multicolumn{1}{l}{} &&\multicolumn{2}{c}{$\Lambda$CDM}&& \multicolumn{3}{c}{$w$CDM}&& \multicolumn{4}{c}{CPL}
\\
\cline{3-4} \cline{6-8} \cline{10-13} % \cline{1-1}
Data& &$\Omega_{\rm m}/10^{-3}$ &$H_0/10^{-1}$  &  &$\Omega_{\rm m}/10^{-3}$ &$H_0/10^{-1}$  &$w/10^{-2}$ &  &$\Omega_{\rm m}/10^{-3}$ &$H_0/10^{-1}$ &$w_0/10^{-2}$ &$w_a/10^{-1}$
\\
\hline
CMB
        &
        & $8.3$
        & $5.9$
        &
        & $34$
        & N/A
        & $25$
        &
        & $41$
        & N/A
        & $48$
        & N/A
        \\
CMB+BAO
        &
        & $6.1$
        & $4.5$
        &
        & $12$
        & $15$
        & $5.9$
        &
        & $25$
        & $24$
        & $26$
        & $7.5$
        \\

CBS
        &
        & $6.0$
        & $4.4$
        &
        & $7.6$
        & $8.2$
        & $3.3$
        &
        & $7.8$
        & $8.3$
        & $8.2$
        & $3.2$
        \\
Tianlai
        &
        & $2.7$
        & $2.0$
        &
        & $3.2$
        & $4.3$
        & $2.4$
        &
        & $20$
        & $16$
        & $16$
        & $4.7$
        \\
CBS+Tianlai
        &
        & $2.3$
        & $1.7$
        &
        & $2.5$
        & $2.6$
        & $1.3$
        &
        & $6.6$
        & $6.0$
        & $5.5$
        & $1.6$
        \\
\bottomrule[1pt]
\end{tabular}
\end{table*}

%As shown in Table~\ref{tab:CBS-result}, using Planck 2018 TT,TE,EE+lowE alone is considerably less constraining and allows for large $H_0$ values in the parameter space of $w$CDM, as well as large $H_0$ and small $w_a$ values in CPL.

It is well known that the CMB-alone data (here, the Planck 2018 TT,TE,EE+lowE power spectrum data), as an early-universe probe, cannot tightly constrain the late-universe physics effects, such as the EoS of dynamical dark energy. Thus, in a model of dynamical dark energy, some parameters relevant to late-universe physics can only get lower or upper limits when the CMB-alone data are used. For example, as shown in Table~\ref{tab:CBS-result} (see also Table \ref{tab:21cm-result}), the CMB-alone data can only provide lower limits for $H_0$ in $w$CDM and CPL, and an upper limit for $w_a$ in CPL, so it is not applicable for these cases to show 1$\sigma$ errors in the table. Also, as shown in Fig.~\ref{fig:CBS}, the CMB-alone data lead to strong parameter degeneracies in dynamical dark energy models.

However, due to the parameter degeneracies being broken, CMB+BAO gives tighter constraints and CBS provides almost the best constraints on cosmological parameters so far. Compared with CBS, Tianlai offers even tighter constraints in $\Lambda$CDM and $w$CDM, as clearly shown in Fig.~\ref{fig:CBS}. In CPL, although Tianlai is inferior to CBS, its constraints on $w_0$ and $w_a$ are $38\%$ and $37\%$ better than those of CMB+BAO, respectively. To sum up, Tianlai has obvious advantages over the three mainstream observations, so it can provide a powerful probe for exploring the nature of dark energy in the future.

The combination CBS+Tianlai gives the constraint errors $\sigma(\Omega_{\rm m})=0.0023$ and $\sigma(H_0)=0.17\ \rm km\ s^{-1}\ Mpc^{-1}$ in $\Lambda$CDM, $\sigma(w)=0.013$ in $w$CDM, and $\sigma(w_0)=0.055$ and $\sigma(w_a)=0.16$ in CPL. Compared with the CBS data, the constraints on $\Omega_{\rm m}$, $H_0$, $w$, $w_0$, and $w_a$ are improved by $62\%$, $61\%$, $61\%$, $33\%$, and $50\%$, respectively, by including the Tianlai data.

\subsection{Residual foreground contamination}\label{sec3.4}
\begin{figure*}[htbp]
\includegraphics[scale=0.32]{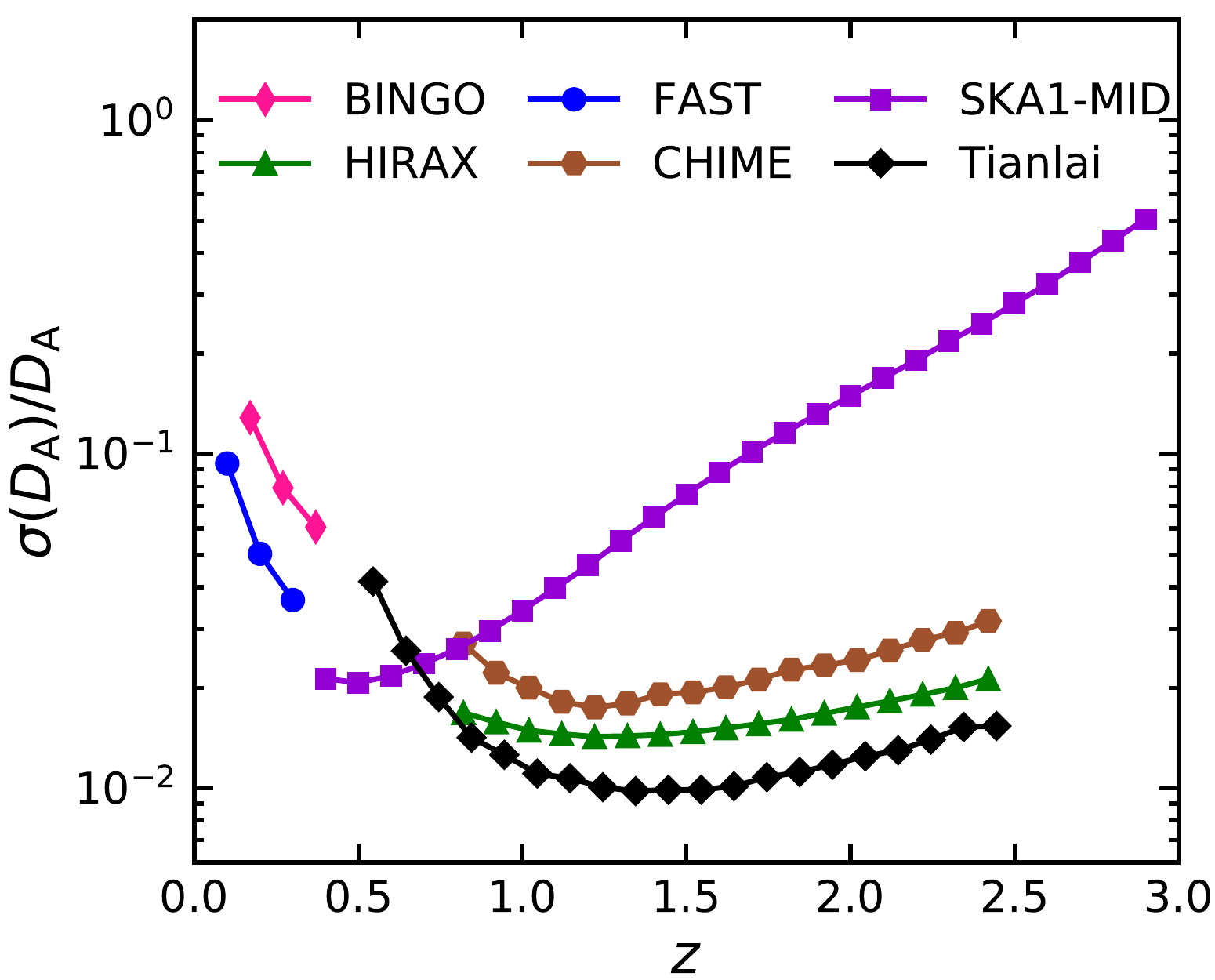}
\includegraphics[scale=0.32]{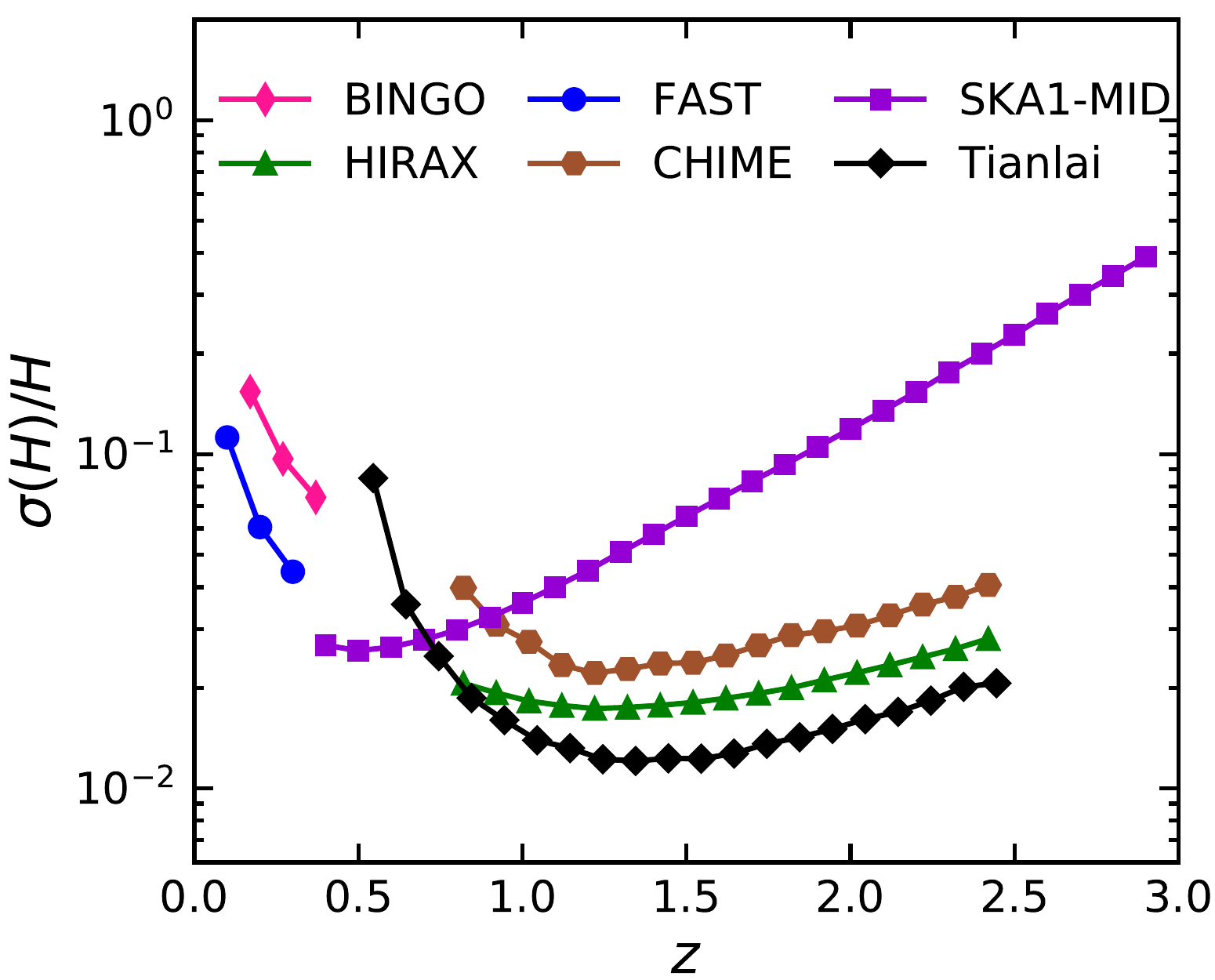}
\includegraphics[scale=0.32]{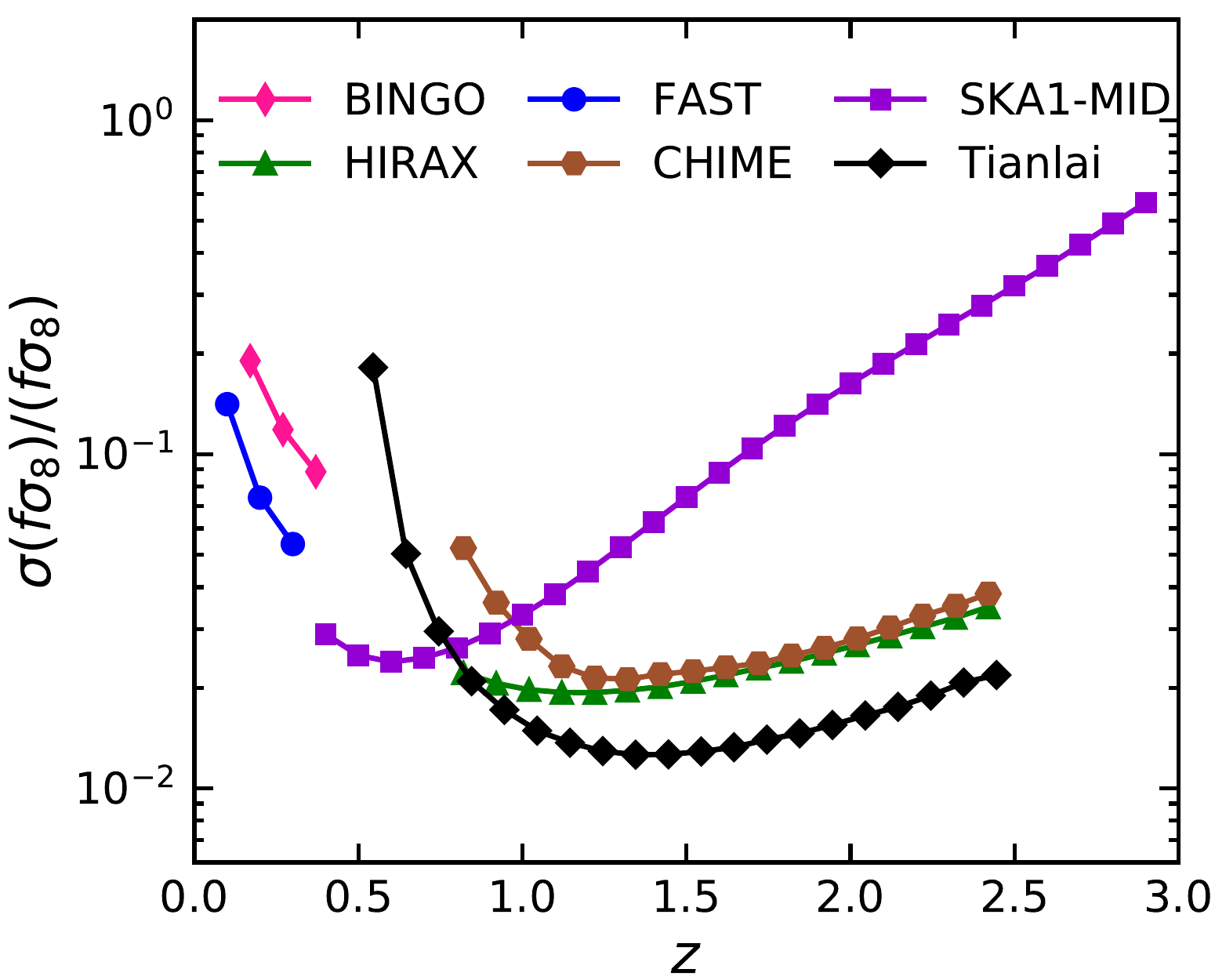}
\centering
\caption{Relative errors on $D_{\rm A}(z)$, $H(z)$, and $[f\sigma_8](z)$ in the case of $\varepsilon_{\rm FG}=10^{-5}$, as a function of redshift.}
\label{fig:frac err 5}
\end{figure*}

\begin{figure}[!htbp]
\centering
\includegraphics[scale=0.43]{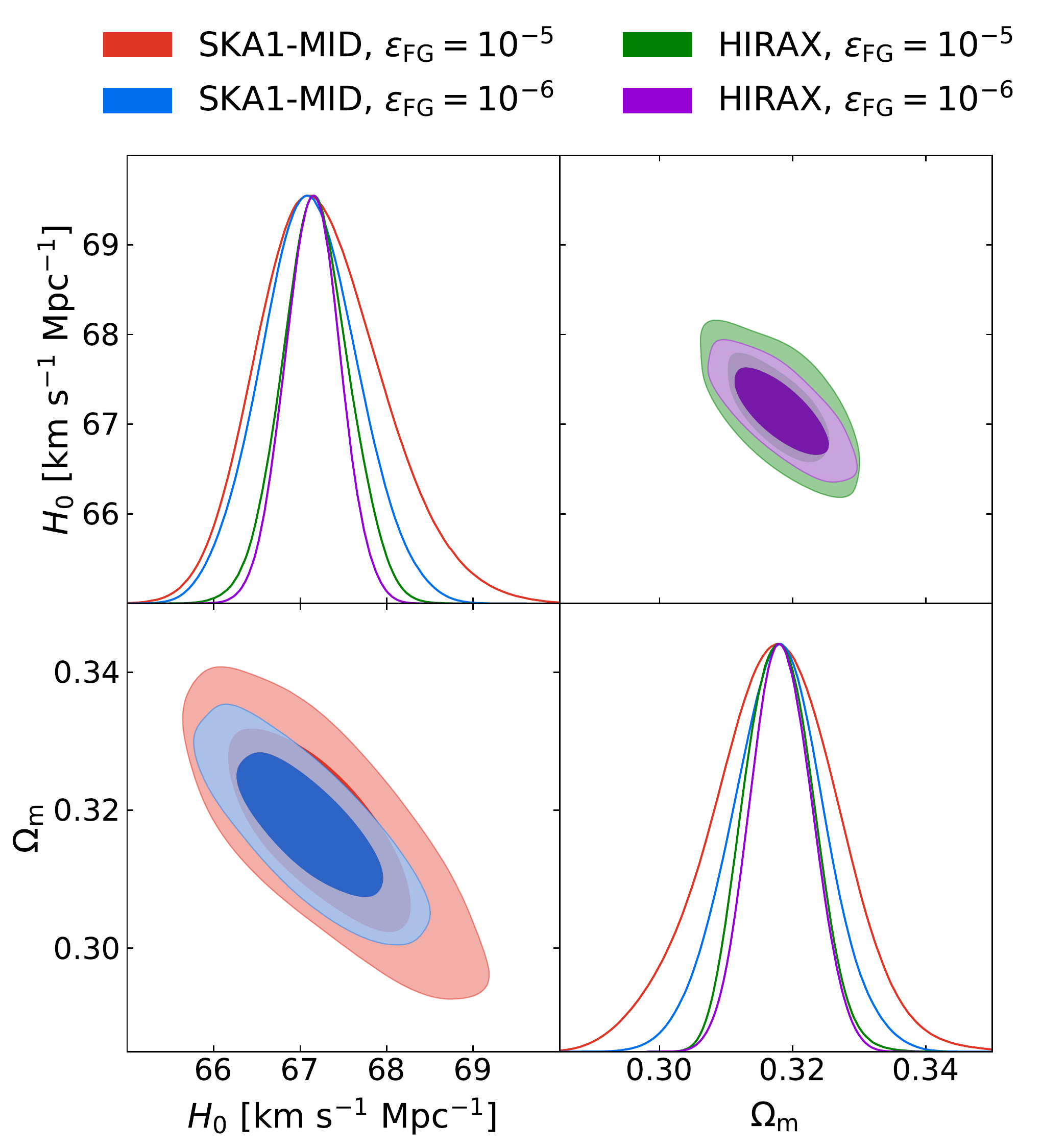}
\caption{Constraints (68.3\% and 95.4\% confidence level) on the $\Lambda$CDM model by using SKA1-MID and HIRAX, in the case of $\varepsilon_{\rm FG}=10^{-6}$ and $\varepsilon_{\rm FG}=10^{-5}$, respectively.}
\label{fig:FG-LCDM}
\end{figure}

\begin{figure}[!htbp]
\includegraphics[scale=0.43]{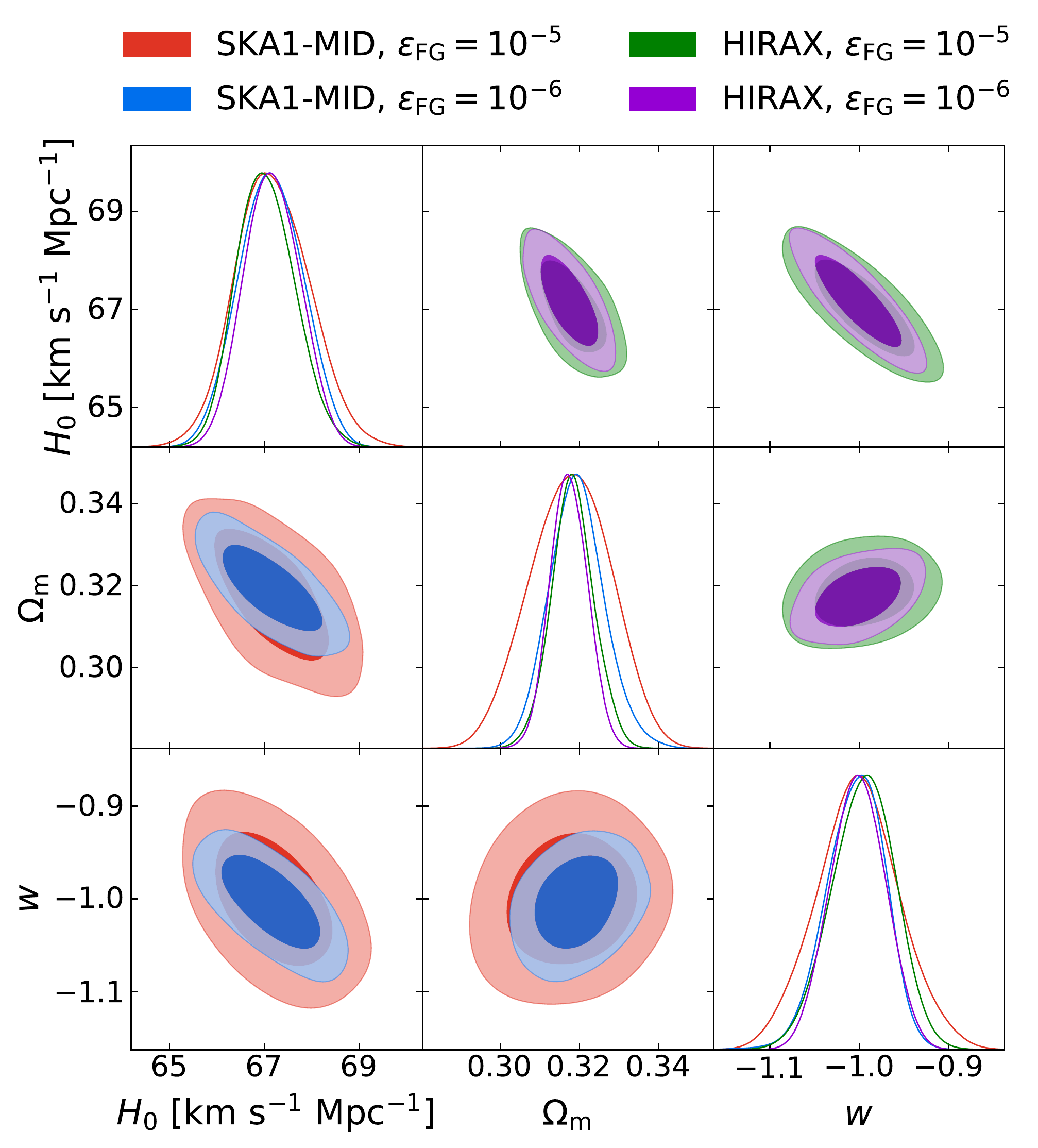}
\centering
\caption{Same as Fig. \ref{fig:FG-LCDM} but for the $w$CDM model.}
\label{fig:FG-wCDM}
\end{figure}

\begin{figure}[!htbp]
\includegraphics[scale=0.43]{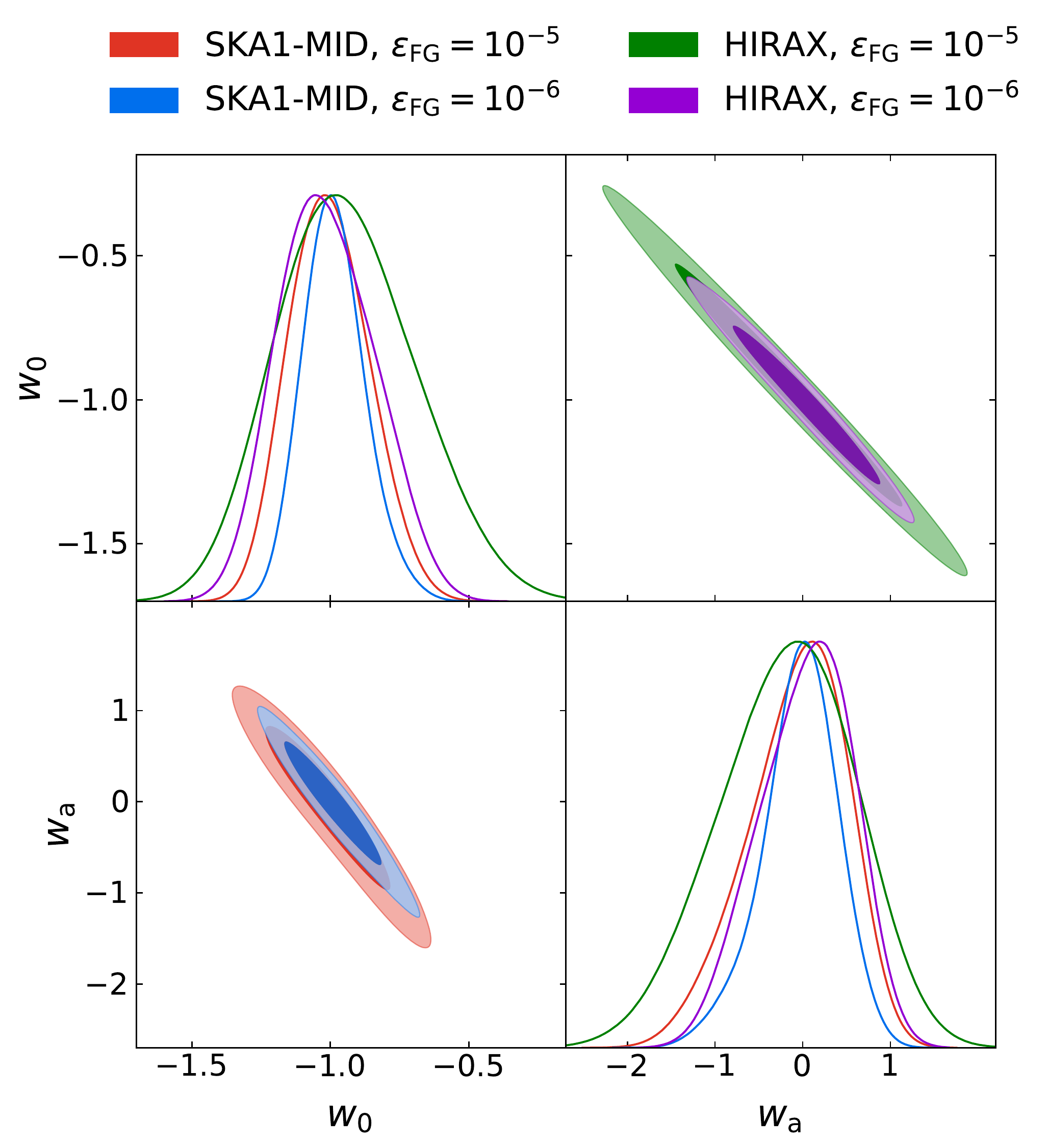}
\centering
\caption{Same as Fig. \ref{fig:FG-LCDM} but for the CPL model.}
\label{fig:FG-CPL}
\end{figure}

In this subsection, we turn to study the effect of residual foreground on constraint results. In the case of $\varepsilon_{\rm FG}=10^{-5}$, we re-simulate the 21 cm IM data and use them to constrain the dark energy models. The forecasted relative errors on $D_{\rm A}(z)$, $H(z)$ and $[f\sigma_8](z)$ are shown in Fig.~\ref{fig:frac err 5}. Notably, the errors given by the SKA1-MID data are very sensitive to the residual foreground contamination amplitude, even reaching $40\%$ at $z\sim3$. In contrast, other experiments are less affected, mainly for the following reasons. First, SKA1-MID is used as a collection of single dishes for IM, so it has relatively low angular resolution at $z\gtrsim 1$ compared to the arrays in interferometric mode in the IM survey, HIRAX, CHIME and Tianlai, leaving it not sensitive to the transverse BAO feature \citep{Villaescusa-Navarro:2016kbz, Ansari:2018ury}. Unfortunately, residual foreground mainly affects the signal extraction on large scales (as shown in Fig.~\ref{fig:CF-FG6}), so SKA1-MID is most affected when the residual foreground contamination is enhanced. Second, BINGO and FAST only detect the low-redshift signals, which are relatively strong and not easy to be dwarfed by the residual foreground. Actually, even though the residual foreground contamination level reaches $\varepsilon_{\rm FG}=10^{-5}$, it is still relatively small, and as we can see, the differences for the cosmological observables between the cases of $\varepsilon_{\rm FG}=10^{-6}$ and $10^{-5}$ (as shown in Fig.~\ref{fig:frac err 6} and Fig.~\ref{fig:frac err 5}) are very marginal (except SKA1-MID). Here we choose the dish arrays, SKA1-MID and HIRAX, as examples to illustrate the effect of residual foreground on cosmological constraints. The results are shown in Figs. \ref{fig:FG-LCDM}--\ref{fig:FG-CPL} and Table~\ref{tab:FG-result}.

We can see that the increased residual foreground weakens the constraints on cosmological parameters. In the $\Lambda$CDM and $w$CDM models, SKA1-MID is more affected than HIRAX, as shown in Figs.~\ref{fig:FG-LCDM}--\ref{fig:FG-wCDM}. Quantitatively, in $\Lambda$CDM, the constraint precision of $\Omega_{\rm m}$ and $H_0$ drop by $36\%$ and $27\%$ for SKA1-MID, and $8.7\%$ and $22\%$ for HIRAX, respectively. In $w$CDM, the constraint accuracy of $w$ drops by $44\%$ and $20\%$, respectively. However, in the CPL model, HIRAX is more affected, as shown in Fig.~\ref{fig:FG-CPL}. The constraint precision of $w_0$ and $w_a$ decrease by $27\%$ and $36\%$ for SKA1-MID, and $39\%$ and $46\%$ for HIRAX, respectively. We have checked that CHIME and Tianlai have similar behaviors to HIRAX. Actually, the performance of SKA1-MID at $0.35<z<0.77$ is not significantly affected, as shown in Fig.~\ref{fig:frac err 5}, so it still has the advantage over the interferometers in constraining the dynamical dark-energy EoS parameters. To sum up, although the cosmological signal can be extracted when $\varepsilon_{\rm FG}\lesssim10^{-5}$, the impact of $\varepsilon_{\rm FG}$ changing from $10^{-6}$ to $10^{-5}$ is very considerable.

In recent years, most forecasts for cosmological parameter estimation with 21 cm IM experiments have assumed perfect foreground removal (e.g., Refs.~\citep{Xu:2014bya,Bull:2015esa,Pourtsidou:2016dzn,Xu:2017rfo,Yohana:2019ahg,Zhang:2021yof,Jin:2021pcv,Xiao:2021nmk}), with exceptions (e.g., Refs.~\citep{Bull:2014rha,Smoot:2014oia,Karagiannis:2019jjx}). As we have shown, the constraint precision depend largely on the residual foreground contamination amplitude. Therefore, on the basis of being able to extract the cosmological signal, we must find ways to further suppress the foreground, so that we can measure the cosmological parameters more accurately.

\begin{table*}[htbp]
\caption{The 1$\sigma$ errors on the cosmological parameters in the $\Lambda$CDM, $w$CDM, and CPL models, by using SKA1-MID and HIRAX, in the case of $\varepsilon_{\rm FG}=10^{-6}$ and $\varepsilon_{\rm FG}=10^{-5}$, respectively. Here, $H_0$ is in units of $\rm km\ s^{-1}\ Mpc^{-1}$.}
\label{tab:FG-result}
\footnotesize
\setlength\tabcolsep{1.5pt}
\renewcommand{\arraystretch}{1.5}
\centering
\begin{tabular}{lccccccccccccccc}
\toprule[1pt]
\multicolumn{1}{l}{} & &&\multicolumn{2}{c}{$\Lambda$CDM}&& \multicolumn{3}{c}{$w$CDM}&& \multicolumn{4}{c}{CPL}
\\
\cline{4-5} \cline{7-9} \cline{11-14} % \cline{1-2}
Data &$\varepsilon_{\rm FG}$& &$\Omega_{\rm m}/10^{-3}$ &$H_0/10^{-1}$  &  &$\Omega_{\rm m}/10^{-3}$ &$H_0/10^{-1}$  &$w/10^{-2}$ &  &$\Omega_{\rm m}/10^{-3}$ &$H_0/10^{-1}$ &$w_0/10^{-2}$ &$w_a/10^{-1}$
\\
\hline
SKA1-MID
        & $10^{-6}$
        &
        & $7.0$
        & $5.5$
        &
        & $6.8$
        & $6.6$
        & $3.2$
        &
        & $15$
        & $11$
        & $11$
        & $4.2$
        \\
SKA1-MID
        & $10^{-5}$
        &
        & $9.5$
        & $7.0$
        &
        & $10$
        & $7.8$
        & $4.6$
        &
        & $19$
        & $12$
        & $14$
        & $5.7$
        \\
HIRAX
        & $10^{-6}$
        &
        & $4.6$
        & $3.2$
        &
        & $4.7$
        & $5.8$
        & $3.0$
        &
        & $22$
        & $18$
        & $18$
        & $5.4$
        \\
HIRAX
        & $10^{-5}$
        &
        & $5.0$
        & $3.9$
        &
        & $5.3$
        & $6.1$
        & $3.6$
        &
        & $32$
        & $24$
        & $25$
        & $7.9$
        \\
\bottomrule[1pt]
\end{tabular}
\end{table*}

\section{Conclusion}\label{sec4}
In this work, we explore the role of future 21 cm intensity mapping experiments, including BINGO, FAST, SKA1-MID, HIRAX, CHIME, and Tianlai, in constraining cosmological parameters. We use the full 21 cm IM power spectrum to constrain the angular diameter distance, expansion rate, and structure growth rate, and then use them to constrain cosmological parameters. In the process of simulating the 21 cm IM data, we consider the influence of residual foreground. We take three most typical dark energy models as examples to complete our analysis, i.e., the $\Lambda$CDM, $w$CDM, and CPL models.

Among the six experiments, BINGO and FAST are relatively weak in cosmological parameter estimation due to their narrow redshift ranges. Due to larger survey area and aperture size, as well as lower receiver noise, FAST gives tighter constraints than BINGO. The high-resolution interferometers, HIRAX, CHIME, and Tianlai, have great advantages in constraining cosmological parameters, especially the Tianlai cylinder array, which alone achieves precision cosmology in $\Lambda$CDM and provides the tightest constraints in $w$CDM. SKA1-MID performs very well in constraining dynamical dark energy models, and offers the best constraints for CPL. Taking full advantages of relevant experiments, we propose a novel survey strategy, specifically, FAST ($0<z<0.35$) + SKA1-MID ($0.35<z<0.77$) + Tianlai ($0.77<z<2.55$). Here, Tianlai can be replaced by CHIME or HIRAX. We plan to conduct more studies on this strategy in future works.

We show that the 21 cm IM data can effectively break the parameter degeneracies inherent in the CMB data. For example, in $\Lambda$CDM, CMB+Tianlai improves the constraints on $\Omega_{\rm m}$ and $H_0$ by $71\%$ and $71\%$, respectively, compared with CMB alone. We find that combining with CMB can highlight the advantage of SKA1-MID in constraining the dynamical dark-energy EoS parameters. Notably, CMB+SKA1-MID can constrain $w$ to $\sigma(w)=0.013$, which is very close to the level of precision cosmology, and can constrain $w_0$ and $w_a$ to $\sigma(w_0)=0.080$ and $\sigma(w_a)=0.25$, which are comparable with the CBS results.

We compare Tianlai with CMB, CMB+BAO, and CBS to illustrate the role of 21 cm IM experiments in cosmological parameter constraints. It is find that Tianlai provides even tighter constraints than CBS in $\Lambda$CDM and $w$CDM. In CPL, although Tianlai is inferior to CBS, its constraints on $w_0$ and $w_a$ are $38\%$ and $37\%$ better than those of CMB+BAO, respectively. Moreover, the combination CBS+Tianlai gives exciting constraints, $\sigma(w)=0.013$, $\sigma(w_0)=0.055$, and $\sigma(w_a)=0.13$. Note that, for Tianlai, we have ignored the baselines shorter than $15\,\rm m$, while in the case of no baseline cutting \citep{Xu:2014bya}, Tianlai can provide better constraints.

We investigate the effect of residual foreground on constraint results. In terms of signal quality, the increased residual foreground has a greater impact on SKA1-MID than other experiments. We choose SKA1-MID and HIRAX for comparison. It is found that the constraint precision of SKA1-MID decrease more in $\Lambda$CDM and $w$CDM. However, in CPL, the constraint precision of HIRAX drop more. This is because the performance of SKA1-MID at low redshifts is also less affected, so it still maintains the advantage in constraining the dynamical dark energy models. Importantly, the increased residual foreground weakens the constraints by several tens percent, which shows that extremely accurate foreground removal techniques are necessary.

Our results are sufficient to show that future 21 cm IM experiments will provide a powerful probe for exploring the nature of dark energy. In the future, the proposed Stage \uppercase\expandafter{\romannumeral2} 21 cm experiment \citep{Ansari:2018ury} will be able to survey larger volumes with higher resolution. We plan to investigate its capability of constraining cosmological parameters in future works.

\begin{acknowledgments}

We thank Ming Zhang, Shang-Jie Jin, Yue Shao, Jing-Zhao Qi, Yi-Chao Li, Yidong Xu, and Jing-Fei Zhang for helpful discussions. This work was supported by the National Natural Science Foundation of China (Grants Nos. 11975072, 11835009, 11875102, and 11690021), the Liaoning Revitalization Talents Program (Grant No. XLYC1905011), the Fundamental Research Funds for the Central Universities (Grant No. N2005030), the National Program for Support of Top-Notch Young Professionals (Grant No. W02070050), and the National 111 Project of China (Grant No. B16009).

\end{acknowledgments}

\bibliography{21cmIM}{}

\providecommand{\href}[2]{#2}\begingroup\raggedright\begin{thebibliography}{10}

\bibitem{Perlmutter:1998np}
{\scshape Supernova Cosmology Project} collaboration, \emph{{Measurements of
  $\Omega$ and $\Lambda$ from 42 high redshift supernovae}},
  \href{https://doi.org/10.1086/307221}{\emph{Astrophys. J.} {\bfseries 517}
  (1999) 565} [\href{https://arxiv.org/abs/astro-ph/9812133}{{\ttfamily
  astro-ph/9812133}}].

\bibitem{Riess:1998cb}
{\scshape Supernova Search Team} collaboration, \emph{{Observational evidence
  from supernovae for an accelerating universe and a cosmological constant}},
  \href{https://doi.org/10.1086/300499}{\emph{Astron. J.} {\bfseries 116}
  (1998) 1009} [\href{https://arxiv.org/abs/astro-ph/9805201}{{\ttfamily
  astro-ph/9805201}}].

\bibitem{Weinberg:2013agg}
D.H.~Weinberg, M.J.~Mortonson, D.J.~Eisenstein, C.~Hirata, A.G.~Riess and
  E.~Rozo, \emph{{Observational Probes of Cosmic Acceleration}},
  \href{https://doi.org/10.1016/j.physrep.2013.05.001}{\emph{Phys. Rept.}
  {\bfseries 530} (2013) 87} [\href{https://arxiv.org/abs/1201.2434}{{\ttfamily
  1201.2434}}].

\bibitem{Blake:2003rh}
C.~Blake and K.~Glazebrook, \emph{{Probing dark energy using baryonic
  oscillations in the galaxy power spectrum as a cosmological ruler}},
  \href{https://doi.org/10.1086/376983}{\emph{Astrophys. J.} {\bfseries 594}
  (2003) 665} [\href{https://arxiv.org/abs/astro-ph/0301632}{{\ttfamily
  astro-ph/0301632}}].

\bibitem{Seo:2003pu}
H.-J.~Seo and D.J.~Eisenstein, \emph{{Probing dark energy with baryonic
  acoustic oscillations from future large galaxy redshift surveys}},
  \href{https://doi.org/10.1086/379122}{\emph{Astrophys. J.} {\bfseries 598}
  (2003) 720} [\href{https://arxiv.org/abs/astro-ph/0307460}{{\ttfamily
  astro-ph/0307460}}].

\bibitem{Guzzo:2008ac}
L.~Guzzo et~al., \emph{{A test of the nature of cosmic acceleration using
  galaxy redshift distortions}},
  \href{https://doi.org/10.1038/nature06555}{\emph{Nature} {\bfseries 451}
  (2008) 541} [\href{https://arxiv.org/abs/0802.1944}{{\ttfamily 0802.1944}}].

\bibitem{Wang:2007ht}
Y.~Wang, \emph{{Differentiating dark energy and modified gravity with galaxy
  redshift surveys}},
  \href{https://doi.org/10.1088/1475-7516/2008/05/021}{\emph{JCAP} {\bfseries
  05} (2008) 021} [\href{https://arxiv.org/abs/0710.3885}{{\ttfamily
  0710.3885}}].

\bibitem{Beutler:2013yhm}
{\scshape BOSS} collaboration, \emph{{The clustering of galaxies in the
  SDSS-III Baryon Oscillation Spectroscopic Survey: Testing gravity with
  redshift-space distortions using the power spectrum multipoles}},
  \href{https://doi.org/10.1093/mnras/stu1051}{\emph{Mon. Not. Roy. Astron.
  Soc.} {\bfseries 443} (2014) 1065}
  [\href{https://arxiv.org/abs/1312.4611}{{\ttfamily 1312.4611}}].

\bibitem{Samushia:2013yga}
L.~Samushia et~al., \emph{{The clustering of galaxies in the SDSS-III Baryon
  Oscillation Spectroscopic Survey: measuring growth rate and geometry with
  anisotropic clustering}},
  \href{https://doi.org/10.1093/mnras/stu197}{\emph{Mon. Not. Roy. Astron.
  Soc.} {\bfseries 439} (2014) 3504}
  [\href{https://arxiv.org/abs/1312.4899}{{\ttfamily 1312.4899}}].

\bibitem{Li:2015poa}
Y.-H.~Li, J.-F.~Zhang and X.~Zhang, \emph{{Probing $f(R)$ cosmology with
  sterile neutrinos via measurements of scale-dependent growth rate of
  structure}},
  \href{https://doi.org/10.1016/j.physletb.2015.03.063}{\emph{Phys. Lett. B}
  {\bfseries 744} (2015) 213}
  [\href{https://arxiv.org/abs/1502.01136}{{\ttfamily 1502.01136}}].

\bibitem{Zhao:2017jma}
M.-M.~Zhao, J.-F.~Zhang and X.~Zhang, \emph{{Measuring growth index in a
  universe with massive neutrinos: A revisit of the general relativity test
  with the latest observations}},
  \href{https://doi.org/10.1016/j.physletb.2018.02.042}{\emph{Phys. Lett. B}
  {\bfseries 779} (2018) 473}
  [\href{https://arxiv.org/abs/1710.02391}{{\ttfamily 1710.02391}}].

\bibitem{Beutler:2011hx}
F.~Beutler, C.~Blake, M.~Colless, D.~Jones, L.~Staveley-Smith, L.~Campbell
  et~al., \emph{{The 6dF Galaxy Survey: Baryon Acoustic Oscillations and the
  Local Hubble Constant}},
  \href{https://doi.org/10.1111/j.1365-2966.2011.19250.x}{\emph{Mon. Not. Roy.
  Astron. Soc.} {\bfseries 416} (2011) 3017}
  [\href{https://arxiv.org/abs/1106.3366}{{\ttfamily 1106.3366}}].

\bibitem{Anderson:2013zyy}
{\scshape BOSS} collaboration, \emph{{The clustering of galaxies in the
  SDSS-III Baryon Oscillation Spectroscopic Survey: baryon acoustic
  oscillations in the Data Releases 10 and 11 Galaxy samples}},
  \href{https://doi.org/10.1093/mnras/stu523}{\emph{Mon. Not. Roy. Astron.
  Soc.} {\bfseries 441} (2014) 24}
  [\href{https://arxiv.org/abs/1312.4877}{{\ttfamily 1312.4877}}].

\bibitem{Delubac:2014aqe}
{\scshape BOSS} collaboration, \emph{{Baryon acoustic oscillations in the
  Ly\ensuremath{\alpha} forest of BOSS DR11 quasars}},
  \href{https://doi.org/10.1051/0004-6361/201423969}{\emph{Astron. Astrophys.}
  {\bfseries 574} (2015) A59}
  [\href{https://arxiv.org/abs/1404.1801}{{\ttfamily 1404.1801}}].

\bibitem{Ross:2014qpa}
A.J.~Ross, L.~Samushia, C.~Howlett, W.J.~Percival, A.~Burden and M.~Manera,
  \emph{{The clustering of the SDSS DR7 main Galaxy sample \textendash{} I. A 4
  per cent distance measure at $z = 0.15$}},
  \href{https://doi.org/10.1093/mnras/stv154}{\emph{Mon. Not. Roy. Astron.
  Soc.} {\bfseries 449} (2015) 835}
  [\href{https://arxiv.org/abs/1409.3242}{{\ttfamily 1409.3242}}].

\bibitem{Alam:2016hwk}
{\scshape BOSS} collaboration, \emph{{The clustering of galaxies in the
  completed SDSS-III Baryon Oscillation Spectroscopic Survey: cosmological
  analysis of the DR12 galaxy sample}},
  \href{https://doi.org/10.1093/mnras/stx721}{\emph{Mon. Not. Roy. Astron.
  Soc.} {\bfseries 470} (2017) 2617}
  [\href{https://arxiv.org/abs/1607.03155}{{\ttfamily 1607.03155}}].

\bibitem{McQuinn:2005hk}
M.~McQuinn, O.~Zahn, M.~Zaldarriaga, L.~Hernquist and S.R.~Furlanetto,
  \emph{{Cosmological parameter estimation using 21 cm radiation from the epoch
  of reionization}}, \href{https://doi.org/10.1086/505167}{\emph{Astrophys. J.}
  {\bfseries 653} (2006) 815}
  [\href{https://arxiv.org/abs/astro-ph/0512263}{{\ttfamily
  astro-ph/0512263}}].

\bibitem{Loeb:2008hg}
A.~Loeb and S.~Wyithe, \emph{{Precise Measurement of the Cosmological Power
  Spectrum With a Dedicated 21cm Survey After Reionization}},
  \href{https://doi.org/10.1103/PhysRevLett.100.161301}{\emph{Phys. Rev. Lett.}
  {\bfseries 100} (2008) 161301}
  [\href{https://arxiv.org/abs/0801.1677}{{\ttfamily 0801.1677}}].

\bibitem{Mao:2008ug}
Y.~Mao, M.~Tegmark, M.~McQuinn, M.~Zaldarriaga and O.~Zahn, \emph{{How
  accurately can 21 cm tomography constrain cosmology?}},
  \href{https://doi.org/10.1103/PhysRevD.78.023529}{\emph{Phys. Rev. D}
  {\bfseries 78} (2008) 023529}
  [\href{https://arxiv.org/abs/0802.1710}{{\ttfamily 0802.1710}}].

\bibitem{Lidz:2011dx}
A.~Lidz, S.R.~Furlanetto, S.P.~Oh, J.~Aguirre, T.-C.~Chang, O.~Dore et~al.,
  \emph{{Intensity Mapping with Carbon Monoxide Emission Lines and the
  Redshifted 21 cm Line}},
  \href{https://doi.org/10.1088/0004-637X/741/2/70}{\emph{Astrophys. J.}
  {\bfseries 741} (2011) 70} [\href{https://arxiv.org/abs/1104.4800}{{\ttfamily
  1104.4800}}].

\bibitem{Battye:2012tg}
R.A.~Battye, I.W.A.~Browne, C.~Dickinson, G.~Heron, B.~Maffei and
  A.~Pourtsidou, \emph{{HI intensity mapping : a single dish approach}},
  \href{https://doi.org/10.1093/mnras/stt1082}{\emph{Mon. Not. Roy. Astron.
  Soc.} {\bfseries 434} (2013) 1239}
  [\href{https://arxiv.org/abs/1209.0343}{{\ttfamily 1209.0343}}].

\bibitem{Xu:2014bya}
Y.~Xu, X.~Wang and X.~Chen, \emph{{Forecasts on the Dark Energy and Primordial
  Non-Gaussianity Observations with the Tianlai Cylinder Array}},
  \href{https://doi.org/10.1088/0004-637X/798/1/40}{\emph{Astrophys. J.}
  {\bfseries 798} (2015) 40} [\href{https://arxiv.org/abs/1410.7794}{{\ttfamily
  1410.7794}}].

\bibitem{Bull:2014rha}
P.~Bull, P.G.~Ferreira, P.~Patel and M.G.~Santos, \emph{{Late-time cosmology
  with 21cm intensity mapping experiments}},
  \href{https://doi.org/10.1088/0004-637X/803/1/21}{\emph{Astrophys. J.}
  {\bfseries 803} (2015) 21} [\href{https://arxiv.org/abs/1405.1452}{{\ttfamily
  1405.1452}}].

\bibitem{Xu:2017rfo}
X.~Xu, Y.-Z.~Ma and A.~Weltman, \emph{{Constraining the interaction between
  dark sectors with future HI intensity mapping observations}},
  \href{https://doi.org/10.1103/PhysRevD.97.083504}{\emph{Phys. Rev. D}
  {\bfseries 97} (2018) 083504}
  [\href{https://arxiv.org/abs/1710.03643}{{\ttfamily 1710.03643}}].

\bibitem{Yohana:2019ahg}
E.~Yohana, Y.-C.~Li and Y.-Z.~Ma, \emph{{Forecasts of cosmological constraints
  from HI intensity mapping with FAST, BINGO \textbackslash{}\& SKA-I}},
  \href{https://arxiv.org/abs/1908.03024}{{\ttfamily 1908.03024}}.

\bibitem{Zhang:2019ipd}
J.-F.~Zhang, B.~Wang and X.~Zhang, \emph{{Forecast for weighing neutrinos in
  cosmology with SKA}},
  \href{https://doi.org/10.1007/s11433-019-1516-y}{\emph{Sci. China Phys. Mech.
  Astron.} {\bfseries 63} (2020) 280411}
  [\href{https://arxiv.org/abs/1907.00179}{{\ttfamily 1907.00179}}].

\bibitem{Tramonte:2020csa}
D.~Tramonte and Y.-Z.~Ma, \emph{{The neutral hydrogen distribution in
  large-scale haloes from 21-cm intensity maps}},
  \href{https://doi.org/10.1093/mnras/staa2727}{\emph{Mon. Not. Roy. Astron.
  Soc.} {\bfseries 498} (2020) 5916}
  [\href{https://arxiv.org/abs/2009.02387}{{\ttfamily 2009.02387}}].

\bibitem{Xu:2020uws}
Y.~Xu and X.~Zhang, \emph{{Cosmological parameter measurement and neutral
  hydrogen 21 cm sky survey with the Square Kilometre Array}},
  \href{https://doi.org/10.1007/s11433-020-1544-3}{\emph{Sci. China Phys. Mech.
  Astron.} {\bfseries 63} (2020) 270431}
  [\href{https://arxiv.org/abs/2002.00572}{{\ttfamily 2002.00572}}].

\bibitem{Chang:2010jp}
T.-C.~Chang, U.-L.~Pen, K.~Bandura and J.B.~Peterson, \emph{{Hydrogen 21-cm
  Intensity Mapping at redshift 0.8}},
  \href{https://doi.org/10.1038/nature09187}{\emph{Nature} {\bfseries 466}
  (2010) 463} [\href{https://arxiv.org/abs/1007.3709}{{\ttfamily 1007.3709}}].

\bibitem{Davis:2000vr}
M.~Davis, J.A.~Newman, S.M.~Faber and A.C.~Phillips, \emph{{The DEEP2 Redshift
  Survey}},  in \emph{{Workshop on Deep Fields}}, 12, 2000,
  \href{https://doi.org/10.1007/10854354_66}{DOI}
  [\href{https://arxiv.org/abs/astro-ph/0012189}{{\ttfamily
  astro-ph/0012189}}].

\bibitem{Masui:2012zc}
K.W.~Masui et~al., \emph{{Measurement of 21 cm brightness fluctuations at z
  \textasciitilde{} 0.8 in cross-correlation}},
  \href{https://doi.org/10.1088/2041-8205/763/1/L20}{\emph{Astrophys. J. Lett.}
  {\bfseries 763} (2013) L20}
  [\href{https://arxiv.org/abs/1208.0331}{{\ttfamily 1208.0331}}].

\bibitem{Switzer:2013ewa}
E.R.~Switzer et~al., \emph{{Determination of z\textasciitilde{}0.8 neutral
  hydrogen fluctuations using the 21 cm intensity mapping auto-correlation}},
  \href{https://doi.org/10.1093/mnrasl/slt074}{\emph{Mon. Not. Roy. Astron.
  Soc.} {\bfseries 434} (2013) L46}
  [\href{https://arxiv.org/abs/1304.3712}{{\ttfamily 1304.3712}}].

\bibitem{Anderson:2017ert}
C.J.~Anderson et~al., \emph{{Low-amplitude clustering in low-redshift 21-cm
  intensity maps cross-correlated with 2dF galaxy densities}},
  \href{https://doi.org/10.1093/mnras/sty346}{\emph{Mon. Not. Roy. Astron.
  Soc.} {\bfseries 476} (2018) 3382}
  [\href{https://arxiv.org/abs/1710.00424}{{\ttfamily 1710.00424}}].

\bibitem{Dickinson:2014wda}
C.~Dickinson, \emph{{BINGO - A novel method to detect BAOs using a total-power
  radio telescope}},  in \emph{{49th Rencontres de Moriond on Cosmology}},
  pp.~139--142, 2014 [\href{https://arxiv.org/abs/1405.7936}{{\ttfamily
  1405.7936}}].

\bibitem{Nan:2011um}
R.~Nan, D.~Li, C.~Jin, Q.~Wang, L.~Zhu, W.~Zhu et~al., \emph{{The
  Five-Hundred-Meter Aperture Spherical Radio Telescope (FAST) Project}},
  \href{https://doi.org/10.1142/S0218271811019335}{\emph{Int. J. Mod. Phys. D}
  {\bfseries 20} (2011) 989} [\href{https://arxiv.org/abs/1105.3794}{{\ttfamily
  1105.3794}}].

\bibitem{Smoot:2014oia}
G.F.~Smoot and I.~Debono, \emph{{21 cm intensity mapping with the Five hundred
  metre Aperture Spherical Telescope}},
  \href{https://doi.org/10.1051/0004-6361/201526794}{\emph{Astron. Astrophys.}
  {\bfseries 597} (2017) A136}
  [\href{https://arxiv.org/abs/1407.3583}{{\ttfamily 1407.3583}}].

\bibitem{Santos:2015gra}
M.G.~Santos et~al., \emph{{Cosmology from a SKA HI intensity mapping survey}},
  \href{https://doi.org/10.22323/1.215.0019}{\emph{PoS} {\bfseries AASKA14}
  (2015) 019} [\href{https://arxiv.org/abs/1501.03989}{{\ttfamily
  1501.03989}}].

\bibitem{Braun:2015zta}
R.~Braun, T.~Bourke, J.A.~Green, E.~Keane and J.~Wagg, \emph{{Advancing
  Astrophysics with the Square Kilometre Array}},
  \href{https://doi.org/10.22323/1.215.0174}{\emph{PoS} {\bfseries AASKA14}
  (2015) 174}.

\bibitem{Newburgh:2016mwi}
L.B.~Newburgh et~al., \emph{{HIRAX: A Probe of Dark Energy and Radio
  Transients}}, \href{https://doi.org/10.1117/12.2234286}{\emph{Proc. SPIE Int.
  Soc. Opt. Eng.} {\bfseries 9906} (2016) 99065X}
  [\href{https://arxiv.org/abs/1607.02059}{{\ttfamily 1607.02059}}].

\bibitem{Newburgh:2014toa}
L.B.~Newburgh et~al., \emph{{Calibrating CHIME, A New Radio Interferometer to
  Probe Dark Energy}}, \href{https://doi.org/10.1117/12.2056962}{\emph{Proc.
  SPIE Int. Soc. Opt. Eng.} {\bfseries 9145} (2014) 4V}
  [\href{https://arxiv.org/abs/1406.2267}{{\ttfamily 1406.2267}}].

\bibitem{Bandura:2014gwa}
K.~Bandura et~al., \emph{{Canadian Hydrogen Intensity Mapping Experiment
  (CHIME) Pathfinder}}, \href{https://doi.org/10.1117/12.2054950}{\emph{Proc.
  SPIE Int. Soc. Opt. Eng.} {\bfseries 9145} (2014) 22}
  [\href{https://arxiv.org/abs/1406.2288}{{\ttfamily 1406.2288}}].

\bibitem{2011SSPMA..41.1358C}
X.~{Chen}, \emph{{Radio detection of dark energy{\textemdash}the Tianlai
  project}}, \href{https://doi.org/10.1360/132011-972}{\emph{Scientia Sinica
  Physica, Mechanica \& Astronomica} {\bfseries 41} (2011) 1358}.

\bibitem{2012IJMPS..12..256C}
X.~{Chen}, \emph{{The Tianlai Project: a 21CM Cosmology Experiment}},  in
  \emph{International Journal of Modern Physics Conference Series}, vol.~12 of
  \emph{International Journal of Modern Physics Conference Series},
  pp.~256--263, Mar., 2012,
  \href{https://doi.org/10.1142/S2010194512006459}{DOI}
  [\href{https://arxiv.org/abs/1212.6278}{{\ttfamily 1212.6278}}].

\bibitem{Brax:2012cr}
P.~Brax, S.~Clesse and A.-C.~Davis, \emph{{Signatures of Modified Gravity on
  the 21-cm Power Spectrum at Reionisation}},
  \href{https://doi.org/10.1088/1475-7516/2013/01/003}{\emph{JCAP} {\bfseries
  01} (2013) 003} [\href{https://arxiv.org/abs/1207.1273}{{\ttfamily
  1207.1273}}].

\bibitem{Hall:2012wd}
A.~Hall, C.~Bonvin and A.~Challinor, \emph{{Testing General Relativity with
  21-cm intensity mapping}},
  \href{https://doi.org/10.1103/PhysRevD.87.064026}{\emph{Phys. Rev. D}
  {\bfseries 87} (2013) 064026}
  [\href{https://arxiv.org/abs/1212.0728}{{\ttfamily 1212.0728}}].

\bibitem{Heneka:2018ins}
C.~Heneka and L.~Amendola, \emph{{General Modified Gravity With 21cm Intensity
  Mapping: Simulations and Forecast}},
  \href{https://doi.org/10.1088/1475-7516/2018/10/004}{\emph{JCAP} {\bfseries
  10} (2018) 004} [\href{https://arxiv.org/abs/1805.03629}{{\ttfamily
  1805.03629}}].

\bibitem{Alonso:2014sna}
D.~Alonso, P.G.~Ferreira and M.G.~Santos, \emph{{Fast simulations for intensity
  mapping experiments}},
  \href{https://doi.org/10.1093/mnras/stu1666}{\emph{Mon. Not. Roy. Astron.
  Soc.} {\bfseries 444} (2014) 3183}
  [\href{https://arxiv.org/abs/1405.1751}{{\ttfamily 1405.1751}}].

\bibitem{Wang:2005zj}
X.-M.~Wang, M.~Tegmark, M.~Santos and L.~Knox, \emph{{Twenty-one centimeter
  tomography with foregrounds}},
  \href{https://doi.org/10.1086/506597}{\emph{Astrophys. J.} {\bfseries 650}
  (2006) 529} [\href{https://arxiv.org/abs/astro-ph/0501081}{{\ttfamily
  astro-ph/0501081}}].

\bibitem{Morales:2012kf}
M.F.~Morales, B.~Hazelton, I.~Sullivan and A.~Beardsley, \emph{{Four
  Fundamental Foreground Power Spectrum Shapes for 21 cm Cosmology
  Observations}},
  \href{https://doi.org/10.1088/0004-637X/752/2/137}{\emph{Astrophys. J.}
  {\bfseries 752} (2012) 137}
  [\href{https://arxiv.org/abs/1202.3830}{{\ttfamily 1202.3830}}].

\bibitem{Parsons:2012qh}
A.R.~Parsons, J.C.~Pober, J.E.~Aguirre, C.L.~Carilli, D.C.~Jacobs and
  D.F.~Moore, \emph{{A Per-Baseline, Delay-Spectrum Technique for Accessing the
  21cm Cosmic Reionization Signature}},
  \href{https://doi.org/10.1088/0004-637X/756/2/165}{\emph{Astrophys. J.}
  {\bfseries 756} (2012) 165}
  [\href{https://arxiv.org/abs/1204.4749}{{\ttfamily 1204.4749}}].

\bibitem{Liu:2014bba}
A.~Liu, A.R.~Parsons and C.M.~Trott, \emph{{Epoch of reionization window. I.
  Mathematical formalism}},
  \href{https://doi.org/10.1103/PhysRevD.90.023018}{\emph{Phys. Rev. D}
  {\bfseries 90} (2014) 023018}
  [\href{https://arxiv.org/abs/1404.2596}{{\ttfamily 1404.2596}}].

\bibitem{Shaw:2014khi}
J.R.~Shaw, K.~Sigurdson, M.~Sitwell, A.~Stebbins and U.-L.~Pen, \emph{{Coaxing
  cosmic 21 cm fluctuations from the polarized sky using m-mode analysis}},
  \href{https://doi.org/10.1103/PhysRevD.91.083514}{\emph{Phys. Rev. D}
  {\bfseries 91} (2015) 083514}
  [\href{https://arxiv.org/abs/1401.2095}{{\ttfamily 1401.2095}}].

\bibitem{Zhu:2016esh}
H.-M.~Zhu, U.-L.~Pen, Y.~Yu and X.~Chen, \emph{{Recovering lost 21 cm radial
  modes via cosmic tidal reconstruction}},
  \href{https://doi.org/10.1103/PhysRevD.98.043511}{\emph{Phys. Rev. D}
  {\bfseries 98} (2018) 043511}
  [\href{https://arxiv.org/abs/1610.07062}{{\ttfamily 1610.07062}}].

\bibitem{Zuo:2018gzm}
S.~Zuo, X.~Chen, R.~Ansari and Y.~Lu, \emph{{21 cm Signal Recovery via Robust
  Principal Component Analysis}},
  \href{https://doi.org/10.3847/1538-3881/aaef3b}{\emph{Astron. J.} {\bfseries
  157} (2018) 4} [\href{https://arxiv.org/abs/1801.04082}{{\ttfamily
  1801.04082}}].

\bibitem{Carucci:2020enz}
I.P.~Carucci, M.O.~Irfan and J.~Bobin, \emph{{Recovery of 21 cm intensity maps
  with sparse component separation}},
  \href{https://doi.org/10.1093/mnras/staa2854}{\emph{Mon. Not. Roy. Astron.
  Soc.} {\bfseries 499} (2020) 304}
  [\href{https://arxiv.org/abs/2006.05996}{{\ttfamily 2006.05996}}].

\bibitem{Cunnington:2020njn}
S.~Cunnington, M.O.~Irfan, I.P.~Carucci, A.~Pourtsidou and J.~Bobin,
  \emph{{21-cm foregrounds and polarization leakage: cleaning and mitigation
  strategies}}, \href{https://doi.org/10.1093/mnras/stab856}{\emph{Mon. Not.
  Roy. Astron. Soc.} {\bfseries 504} (2021) 208}
  [\href{https://arxiv.org/abs/2010.02907}{{\ttfamily 2010.02907}}].

\bibitem{Aghanim:2018eyx}
{\scshape Planck} collaboration, \emph{{Planck 2018 results. VI. Cosmological
  parameters}},
  \href{https://doi.org/10.1051/0004-6361/201833910}{\emph{Astron. Astrophys.}
  {\bfseries 641} (2020) A6}
  [\href{https://arxiv.org/abs/1807.06209}{{\ttfamily 1807.06209}}].

\bibitem{Furlanetto:2006jb}
S.~Furlanetto, S.P.~Oh and F.~Briggs, \emph{{Cosmology at Low Frequencies: The
  21 cm Transition and the High-Redshift Universe}},
  \href{https://doi.org/10.1016/j.physrep.2006.08.002}{\emph{Phys. Rept.}
  {\bfseries 433} (2006) 181}
  [\href{https://arxiv.org/abs/astro-ph/0608032}{{\ttfamily
  astro-ph/0608032}}].

\bibitem{Kaiser:1987qv}
N.~Kaiser, \emph{{Clustering in real space and in redshift space}}, {\emph{Mon.
  Not. Roy. Astron. Soc.} {\bfseries 227} (1987) 1}.

\bibitem{Li:2007rpa}
C.~Li, Y.P.~Jing, G.~Kauffmann, G.~Boerner, X.~Kang and L.~Wang,
  \emph{{Luminosity dependence of the spatial and velocity distributions of
  galaxies: Semi-analytic models versus the Sloan Digital Sky Survey}},
  \href{https://doi.org/10.1111/j.1365-2966.2007.11518.x}{\emph{Mon. Not. Roy.
  Astron. Soc.} {\bfseries 376} (2007) 984}
  [\href{https://arxiv.org/abs/astro-ph/0701218}{{\ttfamily
  astro-ph/0701218}}].

\bibitem{Lewis:1999bs}
A.~Lewis, A.~Challinor and A.~Lasenby, \emph{{Efficient computation of CMB
  anisotropies in closed FRW models}},
  \href{https://doi.org/10.1086/309179}{\emph{Astrophys. J.} {\bfseries 538}
  (2000) 473} [\href{https://arxiv.org/abs/astro-ph/9911177}{{\ttfamily
  astro-ph/9911177}}].

\bibitem{Zhang:2021yof}
M.~Zhang, B.~Wang, P.-J.~Wu, J.-Z.~Qi, Y.~Xu, J.-F.~Zhang et~al.,
  \emph{{Prospects for Constraining Interacting Dark Energy Models with 21 cm
  Intensity Mapping Experiments}},
  \href{https://doi.org/10.3847/1538-4357/ac0ef5}{\emph{Astrophys. J.}
  {\bfseries 918} (2021) 56}
  [\href{https://arxiv.org/abs/2102.03979}{{\ttfamily 2102.03979}}].

\bibitem{Bacon:2018dui}
{\scshape SKA} collaboration, \emph{{Cosmology with Phase 1 of the Square
  Kilometre Array: Red Book 2018: Technical specifications and performance
  forecasts}}, \href{https://doi.org/10.1017/pasa.2019.51}{\emph{Publ. Astron.
  Soc. Austral.} {\bfseries 37} (2020) e007}
  [\href{https://arxiv.org/abs/1811.02743}{{\ttfamily 1811.02743}}].

\bibitem{Santos:2004ju}
M.G.~Santos, A.~Cooray and L.~Knox, \emph{{Multifrequency analysis of 21 cm
  fluctuations from the era of reionization}},
  \href{https://doi.org/10.1086/429857}{\emph{Astrophys. J.} {\bfseries 625}
  (2005) 575} [\href{https://arxiv.org/abs/astro-ph/0408515}{{\ttfamily
  astro-ph/0408515}}].

\bibitem{Smith:2002dz}
{\scshape VIRGO Consortium} collaboration, \emph{{Stable clustering, the halo
  model and nonlinear cosmological power spectra}},
  \href{https://doi.org/10.1046/j.1365-8711.2003.06503.x}{\emph{Mon. Not. Roy.
  Astron. Soc.} {\bfseries 341} (2003) 1311}
  [\href{https://arxiv.org/abs/astro-ph/0207664}{{\ttfamily
  astro-ph/0207664}}].

\bibitem{Wuensche:2019znm}
C.A.~Wuensche et~al., \emph{{Baryon acoustic oscillations from Integrated
  Neutral Gas Observations: Broadband corrugated horn construction and
  testing}}, \href{https://doi.org/10.1007/s10686-020-09666-9}{\emph{Exper.
  Astron.} {\bfseries 50} (2020) 125}
  [\href{https://arxiv.org/abs/1911.13188}{{\ttfamily 1911.13188}}].

\bibitem{Wuensche:2021ebn}
C.A.~Wuensche et~al., \emph{{The BINGO Project II: Instrument Description}},
  \href{https://arxiv.org/abs/2107.01634}{{\ttfamily 2107.01634}}.

\bibitem{Ansari:2018ury}
{\scshape Cosmic Visions 21 cm} collaboration, \emph{{Inflation and Early Dark
  Energy with a Stage II Hydrogen Intensity Mapping experiment}},
  \href{https://arxiv.org/abs/1810.09572}{{\ttfamily 1810.09572}}.

\bibitem{Amiri:2019qbv}
{\scshape CHIME/FRB} collaboration, \emph{{Observations of fast radio bursts at
  frequencies down to 400 megahertz}},
  \href{https://doi.org/10.1038/s41586-018-0867-7}{\emph{Nature} {\bfseries
  566} (2019) 230} [\href{https://arxiv.org/abs/1901.04524}{{\ttfamily
  1901.04524}}].

\bibitem{Amiri:2019bjk}
{\scshape CHIME/FRB} collaboration, \emph{{A Second Source of Repeating Fast
  Radio Bursts}},
  \href{https://doi.org/10.1038/s41586-018-0864-x}{\emph{Nature} {\bfseries
  566} (2019) 235} [\href{https://arxiv.org/abs/1901.04525}{{\ttfamily
  1901.04525}}].

\bibitem{CHIMEFRB:2020abu}
{\scshape CHIME/FRB} collaboration, \emph{{A bright millisecond-duration radio
  burst from a Galactic magnetar}},
  \href{https://doi.org/10.1038/s41586-020-2863-y}{\emph{Nature} {\bfseries
  587} (2020) 54} [\href{https://arxiv.org/abs/2005.10324}{{\ttfamily
  2005.10324}}].

\bibitem{CHIMEFRB:2021srp}
{\scshape CHIME/FRB} collaboration, \emph{{The First CHIME/FRB Fast Radio Burst
  Catalog}},  \href{https://arxiv.org/abs/2106.04352}{{\ttfamily 2106.04352}}.

\bibitem{Li:2020ast}
J.~Li et~al., \emph{{The Tianlai Cylinder Pathfinder array: System functions
  and basic performance analysis}},
  \href{https://doi.org/10.1007/s11433-020-1594-8}{\emph{Sci. China Phys. Mech.
  Astron.} {\bfseries 63} (2020) 129862}
  [\href{https://arxiv.org/abs/2006.05605}{{\ttfamily 2006.05605}}].

\bibitem{Witzemann:2017lhi}
A.~Witzemann, P.~Bull, C.~Clarkson, M.G.~Santos, M.~Spinelli and A.~Weltman,
  \emph{{Model-independent curvature determination with 21 cm intensity mapping
  experiments}}, \href{https://doi.org/10.1093/mnrasl/sly062}{\emph{Mon. Not.
  Roy. Astron. Soc.} {\bfseries 477} (2018) L122}
  [\href{https://arxiv.org/abs/1711.02179}{{\ttfamily 1711.02179}}].

\bibitem{Scolnic:2017caz}
D.M.~Scolnic et~al., \emph{{The Complete Light-curve Sample of
  Spectroscopically Confirmed SNe Ia from Pan-STARRS1 and Cosmological
  Constraints from the Combined Pantheon Sample}},
  \href{https://doi.org/10.3847/1538-4357/aab9bb}{\emph{Astrophys. J.}
  {\bfseries 859} (2018) 101}
  [\href{https://arxiv.org/abs/1710.00845}{{\ttfamily 1710.00845}}].

\bibitem{Chevallier:2000qy}
M.~Chevallier and D.~Polarski, \emph{{Accelerating universes with scaling dark
  matter}}, \href{https://doi.org/10.1142/S0218271801000822}{\emph{Int. J. Mod.
  Phys. D} {\bfseries 10} (2001) 213}
  [\href{https://arxiv.org/abs/gr-qc/0009008}{{\ttfamily gr-qc/0009008}}].

\bibitem{Linder:2002et}
E.V.~Linder, \emph{{Exploring the expansion history of the universe}},
  \href{https://doi.org/10.1103/PhysRevLett.90.091301}{\emph{Phys. Rev. Lett.}
  {\bfseries 90} (2003) 091301}
  [\href{https://arxiv.org/abs/astro-ph/0208512}{{\ttfamily
  astro-ph/0208512}}].

\bibitem{Chang:2007xk}
T.-C.~Chang, U.-L.~Pen, J.B.~Peterson and P.~McDonald, \emph{{Baryon Acoustic
  Oscillation Intensity Mapping as a Test of Dark Energy}},
  \href{https://doi.org/10.1103/PhysRevLett.100.091303}{\emph{Phys. Rev. Lett.}
  {\bfseries 100} (2008) 091303}
  [\href{https://arxiv.org/abs/0709.3672}{{\ttfamily 0709.3672}}].

\bibitem{Jin:2021pcv}
S.-J.~Jin, L.-F.~Wang, P.-J.~Wu, J.-F.~Zhang and X.~Zhang, \emph{{How can
  gravitational-wave standard sirens and 21 cm intensity mapping jointly
  provide a precise late-universe cosmological probe?}},
  \href{https://arxiv.org/abs/2106.01859}{{\ttfamily 2106.01859}}.

\bibitem{Villaescusa-Navarro:2016kbz}
F.~Villaescusa-Navarro, D.~Alonso and M.~Viel, \emph{{Baryonic acoustic
  oscillations from 21 cm intensity mapping: the Square Kilometre Array case}},
  \href{https://doi.org/10.1093/mnras/stw3224}{\emph{Mon. Not. Roy. Astron.
  Soc.} {\bfseries 466} (2017) 2736}
  [\href{https://arxiv.org/abs/1609.00019}{{\ttfamily 1609.00019}}].

\bibitem{Bull:2015esa}
P.~Bull, S.~Camera, A.~Raccanelli, C.~Blake, P.~Ferreira, M.~Santos et~al.,
  \emph{{Measuring baryon acoustic oscillations with future SKA surveys}},
  \href{https://doi.org/10.22323/1.215.0024}{\emph{PoS} {\bfseries AASKA14}
  (2015) 024}.

\bibitem{Pourtsidou:2016dzn}
A.~Pourtsidou, D.~Bacon and R.~Crittenden, \emph{{HI and cosmological
  constraints from intensity mapping, optical and CMB surveys}},
  \href{https://doi.org/10.1093/mnras/stx1479}{\emph{Mon. Not. Roy. Astron.
  Soc.} {\bfseries 470} (2017) 4251}
  [\href{https://arxiv.org/abs/1610.04189}{{\ttfamily 1610.04189}}].

\bibitem{Xiao:2021nmk}
L.~Xiao, A.A.~Costa and B.~Wang, \emph{{Forecasts on Interacting Dark Energy
  from 21-cm Angular Power Spectrum with BINGO and SKA observations}},
  \href{https://arxiv.org/abs/2103.01796}{{\ttfamily 2103.01796}}.

\bibitem{Karagiannis:2019jjx}
D.~Karagiannis, A.~Slosar and M.~Liguori, \emph{{Forecasts on Primordial
  non-Gaussianity from 21 cm Intensity Mapping experiments}},
  \href{https://doi.org/10.1088/1475-7516/2020/11/052}{\emph{JCAP} {\bfseries
  11} (2020) 052} [\href{https://arxiv.org/abs/1911.03964}{{\ttfamily
  1911.03964}}].

\end{thebibliography}\endgroup
\bibliographystyle{JHEP}

\end{document}